\documentclass[aip,preprint,floatfix] {revtex4} 
\newcommand{\rvec}{\mathrm {\mathbf {r}}} 
 
\newcommand{\Rvec}{\mathrm {\mathbf {R}}} 
\newcommand{\Cvec}{\mathrm {\mathbf {C}}} 
\newcommand{\Svec}{\mathrm {\mathbf {S}}} 
\newcommand{\Ivec}{\mathrm {\mathbf {I}}}

\newenvironment{rcases}
  {\left.\begin{aligned}}
  {\end{aligned}\right\rbrace}
  
\usepackage{graphicx}
\usepackage{subfigure}
\usepackage{xcolor}
\usepackage{enumerate}
\usepackage{braket}
\usepackage{amsmath, bm}
\usepackage{stackengine}
\usepackage{adjustbox}

\usepackage{color, soul}
\definecolor{darkblue}{rgb}{0,0,0.5}
\setulcolor{darkblue}

\begin{document}

\title{A self-consistent systematic optimization of range-separated hybrid functionals from first principles}

\author{Abhisek Ghosal}
\author{Amlan K.~Roy}
\altaffiliation{Email: akroy@iiserkol.ac.in, akroy6k@gmail.com.}                                           
\affiliation{Department of Chemical Sciences\\
Indian Institute of Science Education and Research (IISER) Kolkata, \\  
Nadia, Mohanpur-741246, WB, India.}

\begin{abstract}
In this communication, we represent a self-consistent systematic optimization procedure for the development of optimally 
tuned (OT) range-separated hybrid (RSH) functionals from \emph{first principles}. This is an offshoot of our recent 
work, which employed a purely numerical approach for efficient computation of exact exchange contribution in the 
conventional global hybrid functionals through a range-separated (RS) technique. We make use of the size-dependency 
based ansatz i.e., RS parameter, $\gamma$, is a functional of density, $\rho(\rvec)$, of which not much is known. To be 
consistent with this ansatz, a novel procedure is presented that relates the characteristic length of a given 
system (where $\rho(\rvec)$ exponentially decays to zero) with $\gamma$ self-consistently via a simple 
mathematical constraint. In practice, $\gamma_{\mathrm{OT}}$ is obtained through an optimization of total energy as 
follows: $\gamma_{\mathrm{OT}} \equiv \underset{\gamma }{\mathrm{opt}} \ E_{\mathrm{tot},\gamma}$.
It is found that the parameter $\gamma_{\mathrm{OT}}$, estimated as above can show better performance in predicting 
properties (especially from frontier orbital energies) than conventional respective RSH functionals, of a given system. 
We have examined the nature of highest fractionally occupied orbital from exact piece-wise linearity behavior, which 
reveals that this approach is sufficient to maintain this condition. A careful statistical analysis then illustrates the 
viability and suitability of the current approach. All the calculations are done in a Cartesian-grid based 
pseudopotential (G)KS-DFT framework. 

\vspace{5mm}

\end{abstract}
\maketitle

\section{Introduction}
Throughout the past several decades, density functional theory (DFT) \citep{hohenberg64} has been invoked to electronic 
structure calculations across an unusually wide variety of fields, from organic chemistry to condensed matter physics, 
as it allows for an accurate quantum mechanical description at a relatively modest computational cost \citep{cohen11, 
burke12, becke14, jones15}. Practical applications of DFT, to a large extent relies on the solution of Kohn-Sham (KS) 
equation \citep{kohn65} or  generalization (G)KS equation \citep{seidl96}. Within such a framework, the interacting 
many-electron system is mapped onto an effective single-particle one through a local one-body potential, called KS 
potential, $v_{\mathrm{KS}}(\rvec)$, keeping the ground-state density fixed. In principle, the theory is \emph{exact}, 
and has the ability to capture many-body effects completely and uniquely. All the pertinent interactions present in 
$v_{\mathrm{KS}}(\rvec)$, are included into a single additive exchange-correlation (XC) potential, 
$v_{\mathrm{xc}}(\rvec)$. It is the functional derivative of XC energy, $E_{\mathrm{xc}}[\rho(\rvec)]$, which includes 
Pauli, correlation and the subtle kinetic energy effects, depending only on $\rho(\rvec)$. However, although the 
existence and uniqueness of $E_{\mathrm{xc}}[\rho(\rvec)]$ is guaranteed, its exact form still remains elusive.

So in this scenario, $E_{\mathrm{xc}}[\rho(\rvec)]$ needs to be approximated in the so-called density functional 
approximation (DFA) hoping that they are sufficiently accurate to be useful. Therefore, the practical success of 
(G)KS-DFT hinges entirely on the existence of suitable DFAs. The commonly used DFAs can be hierarchically categorized 
as local (spin) density approximation L(S)DA \citep{kohn65} (containing $\rho$ only), generalized gradient 
approximation GGA \citep{becke88a, perdew96} (with addition of gradient of electron density, $\nabla\rho$), and 
meta-GGA \citep{tao03, zhao08} (with inclusion of Laplacian of density and kinetic energy density, $\tau$), the third 
rung of Jacob's ladder \citep{perdew00}. With additional introduction of exact exchange (EEX) energy, one gets the 
hybrid functionals \citep{becke93b, perdew96a}, while incorporation of EEX energy density, $e^{\mathrm{x}}$, leads to 
the hyper functionals \citep{becke05, perdew08}, residing in the fourth rung of the ladder. Now we also have functionals 
that go beyond this (to include virtual orbitals), requiring even higher computational cost, and have also been successfully 
implemented in several quantum chemistry programs \citep{grimme07,zhang09}. 

In general, the DFAs experience certain issues regarding (i) piece-wise linearity (PWL) of total energy in the 
fractional number of particles \citep{perdew92,yang00} (ii) non-cancellation of spurious Coulomb self-repulsion 
energy$-$the so-called self-interaction error (SIE) \citep{perdew81,bao18} and (iii) asymptotically correct behavior of 
XC potential at long range (LR) for finite systems \citep{levy84}. These three points are not equivalent, but are 
connected to each other by a certain extent \citep{kronik20}. These are very important conditions for developing 
advanced density functionals, as each of them allows one to avoid different aspects of spurious electron interaction. 
A very promising area in which the above conditions can be controlled in a satisfactory way is via an \emph{optimal 
tuning} (OT) of the range-separated hybrid (RSH) functionals \citep{kronik18, baer10}. Generally, these are based on 
partitioning the Coulomb interaction kernel into a short range (SR) and a LR part, usually through an RS operator, 
$g(\gamma,\rvec)$, and an RS parameter, $\gamma$, in the form of 
\begin{equation}
\frac{1}{\rvec}=\frac{\tilde{g}(\gamma,\rvec)}{\rvec}+\frac{g(\gamma,\rvec)}{\rvec}, 
\end{equation}
where, $\tilde{g}(\gamma,\rvec)$ represents the complementary RS operator. This was first proposed in 
\citep{leininger97} in the field of multi-reference configuration interaction, keeping in mind that the dynamical 
electron correlation hardly affects the LR interactions due to the rapidly decaying features. Here, $\gamma$ plays a 
pivotal role to adjust the contribution of EEX between SR to LR region for a given RS operator. In literature, these 
two regions are treated separately, depending on the system of interest. In general, the SR region is described using a 
modified inter-electronic distance-dependent local or semi-local DFA and the LR sector by EEX with the 
$g(\gamma,\rvec)/\rvec$ correction, mainly in the finite systems. The XC energy, based on above partitioning scheme 
then can be defined as,
\begin{eqnarray}
E_{\mathrm{xc}} = a^{\mathrm{sr}}_{\mathrm{eex}}E^{\mathrm{sr}}_{\mathrm{eex}}(\gamma)+
(1-a^{\mathrm{sr}}_{\mathrm{eex}})E^{\mathrm{x,sr}}_{\mathrm{dfa}}(\gamma)+
b^{\mathrm{lr}}_{\mathrm{eex}} E^{\mathrm{lr}}_{\mathrm{eex}}(\gamma)+ 
	(1-b^{\mathrm{lr}}_{\mathrm{eex}})E^{\mathrm{x,lr}}_{\mathrm{dfa}}(\gamma)+ 
	E^{\mathrm{c}}_{\mathrm{dfa}},
\end{eqnarray}
where $E^{\mathrm{sr}}_{\mathrm{eex}}$, $E^{\mathrm{lr}}_{\mathrm{eex}}$ refer to EEX energy contribution while 
$E^{\mathrm{x,sr}}_{\mathrm{dfa}}$, $E^{\mathrm{x,lr}}_{\mathrm{dfa}}$ denote DFA exchange, at SR and LR regions 
respectively. A particular set of ($a^{\mathrm{sr}}_{\mathrm{eex}}$, $b^{\mathrm{lr}}_{\mathrm{eex}}$) defines a 
specific mode of partitioning for a given $g(\gamma,\rvec)$. RSH functionals with $b^{\mathrm{lr}}_{\mathrm{eex}}=1$ 
make XC potential asymptotically correct at LR region. At the same time, a delicate balance between EEX and dynamical 
correlation is retained by adopting an optimal value of $a^{\mathrm{sr}}_{\mathrm{eex}}$. Hence, these functionals are 
not fully SIE free. Moreover, they do not follow the PWL condition unless optimally tuned of a given system. This is 
due to the default value of $\gamma$, which is usually obtained semi-empirically by fitting the reference data 
\citep{iikura01, tawada04, yanai04, lange08}. 

In OT procedure, $\gamma$ is usually determined from \emph{first principles} by actively enforcing Koopmans' theorem 
\citep{salzner09}. It satisfies PWL conditions, and in principle, would preserve the size-dependency of $\gamma$ 
on $\rho$, but is very hard to maintain with a universal value of $\gamma$ \citep{baer10}. It makes the XC potential 
asymptotically correct at LR region. This improves properties based on orbital energies, such as vertical ionization 
energy (IE), electron affinity (EA), fundamental gap (FG), optical gap, charge-transfer excitation as well as Rydberg 
excitation \citep{livshits07, stein10}. Besides the traditional $\gamma$-tuning scheme based on satisfying 
Koopmans' theorem, recent advances based on electron localization function and localized orbital locator have been 
developed in the literature which requires only one single self-consistent field calculation \citep{borpuzari17, wang18}. 
Moreover, a self-consistent OT-RSH approach \citep{tamblyn14} has been developed based on the minimization 
of inter-atomic forces, which offers better geometries and vibrational modes. 

These developments motivated us to work further in this area by asking ``how else one can self-consistently and 
systematically optimize $\gamma$ for a given system from \emph{first principles} irrespective of the properties of 
interest". That lies at the heart of constructing \emph{ab initio} OT-RSH functionals. Here, we present such a 
procedure, taking inspiration from our recent work published elsewhere \citep{ghosal19}, where we pursued a fully 
numerical approach for efficient computation of EEX contribution. A size-dependency based ansatz is invoked i.e., 
$\gamma$ is a functional of density. Thus a systematic procedure is presented that relates $\rho(\rvec)$ with $\gamma$ 
via a simple mathematical constraint following the arguments of \citep{ghosal16, ghosal18, ghosal19}. In practice, we 
obtain $\gamma_{\mathrm{OT}}$ through an optimization of total energy as follows: 
$\gamma_{\mathrm{OT}} \equiv \underset{\gamma }{\mathrm{opt}} \ E_{\mathrm{tot},\gamma}$. The suitability, efficacy and 
performance of this $\gamma_{\mathrm{opt}}$ is analyzed by a comparison of certain properties derived from orbital 
energies. Next, we have examined the nature of highest fractionally occupied orbital from exact PWL 
behavior. Moreover, we also consider small chains versus small conjugated molecules to reveal the effect of conjugation 
in determining $\gamma_{\text{OT}}$. At the end, a comparison between present method with the traditional OT  
strategy is discussed.

The manuscript is organized as follows. In next section, we briefly review the general framework of Cartesian 
coordinate grid (CCG) based pseudopotential (G)KS-DFT. Then, we present the theoretical background for SR/LR-EEX 
through Fourier convolution theorem (FCT) in CCG. Next, we discuss in detail about optimal tuning of $\gamma$. In 
Sec.~III, computational details are given. In Sec.~IV, we illustrate the performance of proposed scheme considering 
following properties such as IE, EA, FG and fractional occupation for a decent number of atoms ($15$) and molecules 
(20). These are presented for LC-BLYP, CAM-B3LYP, LC-PBE, CAM-PBE0, LRC-$\omega$PBEh$^{\star}$ functional along with 
their presently described OT version (denoted by ``ot" subscript), as well as global hybrids B3LYP and PBE0 functionals. 
Finally, conclusions as well as future and outlook are summarized in the last section.

\section{Methodology}
\subsection{Pseudopotential (G)KS-DFT in CCG}
For a many-electron system, one can write the single-particle (G)KS equation in presence of pseudopotential as (atomic unit 
employed unless stated otherwise), 
\begin{equation}
\bigg[ -\frac{1}{2} \nabla^2 + v^{\mathrm{p}}_{\mathrm{ion}}(\rvec) + v_{\mathrm{ext}}(\rvec) + v_{\mathrm{H}}[\rho(\rvec)] 
	+ v_{\mathrm{xc}}[\rho(\rvec)] \bigg] \phi_i^{\sigma}(\rvec) = \epsilon_{i}^{\sigma} \phi_i^{\sigma}(\rvec),
\end{equation}
where $v^{\mathrm{p}}_{\mathrm{ion}}$ denotes the ionic pseudopotential, written as below, 
\begin{equation}
v^{\mathrm{p}}_{\mathrm{ion}}(\rvec) = \sum_{\Rvec_a} v^{\mathrm{p}}_{\mathrm{ion},\mathrm{a}} (\rvec-\Rvec_a).
\end{equation}
In the above equation, $v^{\mathrm{p}}_{\mathrm{ion},\mathrm{a}}$ signifies ion-core pseudopotential associated with atom 
A, situated at $\Rvec_a$. The classical Coulomb (Hartree) term, $v_{\mathrm{H}}[\rho(\rvec)]$ describes usual 
electrostatic interaction amongst valence electrons whereas $v_{\mathrm{xc}}[\rho(\rvec)]$ signifies the non-classical XC 
part of latter, and $\{ \phi^{\sigma}_{i},\sigma= \alpha \quad \mathrm{or} \quad  \beta \}$ corresponds to a set of $N$ 
occupied orthonormal spin-MOs. Within LCAO-MO approximation, the coefficients for expansion of spin-MOs satisfy a set of 
equations, very similar to that in HF theory,  
\begin{equation}
\sum_{\nu} F_{\mu \nu}^{\sigma} C_{\nu i}^{\sigma} = \epsilon_{i}^{\sigma} \sum_{\nu}^{\sigma} S_{\mu \nu} 
C_{\nu i}^{\sigma},
\end{equation}
satisfying the orthonormality condition, $(\Cvec^{\sigma})^{\dagger} \Svec \Cvec^{\sigma} = \Ivec$. Here $\Cvec^{\sigma}$ 
contains the respective spin-MO coefficients $\{C_{\nu i}^{\sigma}\}$ for a given spin-MO $\phi_i^{\sigma}(\rvec)$, $\Svec$ 
is the usual overlap matrix corresponding to elements $S_{\mu \nu}$, $\bm{\epsilon}^{\sigma}$ refers to diagonal matrix of 
spin-MO eigenvalues $\{\epsilon_{i}^{\sigma}\}$. The (G)KS-Fock matrix has elements $F_{\mu \nu}^{\sigma}$, constituting of 
following contributions, 
\begin{equation}
F_{\mu \nu}^{\sigma} = H_{\mu \nu}^{\mathrm{core}} + J_{\mu \nu} + F_{\mu \nu}^{\mathrm{xc}\sigma}.
\end{equation} 
In this equation, all one-electron contributions, such as kinetic energy, nuclear-electron attraction and pseudopotential 
matrix elements are included in first term, whereas $J_{\mu \nu}$ and $F_{\mu \nu}^{\mathrm{xc}\sigma}$ account for 
classical Hartree and XC potentials with EEX respectively. 
 
Now we discretize various quantities like localized basis function, electron density, MO as well as two-electron potentials 
directly on a 3D cubic box having $x,y,z$ axes, 
\begin{eqnarray}
r_{i}=r_{0}+(i-1)h_{r}, \quad i=1,2,3,....,N_{r}~, \quad r_{0}=-\frac{N_{r}h_{r}}{2}, \quad  r \in \{ x,y,z \},
\end{eqnarray}
where $h_{r}, N_r$ denote grid spacing and total number of points along each directions respectively. The electron density 
$\rho(\rvec)$ in this grid may be simply written as (``g" symbolizes discretized grid),
\begin{equation}
\rho(\rvec_g) = \sum_{\mu,\nu} P_{\mu \nu } \chi_{\mu}(\rvec_g) \chi_{\nu}(\rvec_g), 
\end{equation}
where $\{\chi_{\mu}\}$ corresponds to atomic orbitals (AOs). At this stage, the two-electron contributions of KS matrix are 
computed through direct numerical integration in the grid, 
\begin{equation}
F_{\mu \nu}^{\mathrm{Hxc}} = \langle \chi_{\mu}(\rvec_g)|v_{\mathrm{Hxc}}(\rvec_g)|\chi_{\nu}(\rvec_g) \rangle = h_x h_y 
h_z \sum_g \chi_{\mu}(\rvec_g) v_{\mathrm{Hxc}}(\rvec_g) \chi_{\nu}(\rvec_g).
\end{equation}
where $v_{Hxc}(\rvec_g)$ refers to the combined Hartree and XC potential. The detailed construction of various potentials 
in CCG has been well documented in our earlier work \citep{roy08, roy08a, roy10, roy11, roy09, ghosal16, ghosal18}; hence 
not repeated here.

\subsection{SR/LR exact exchange through FCT}
A well-defined numerical methodology for the EEX energy and potential was recently developed in our group \citep{ghosal19}. 
Accordingly the SR/LR EEX energy, $E_{\mathrm{eex},\sigma}^{\mathrm{sr/lr}}$, 
can be computed numerically by integrating the corresponding density $e^{\mathrm{sr/lr}}_{\mathrm{eex},\sigma}(\rvec)$ as 
given by, 
\begin{equation}
	E^{\mathrm{sr/lr}}_{\mathrm{eex},\sigma}=\frac{1}{2}\int e^{\mathrm{sr/lr}}_{\mathrm{eex},\sigma}(\rvec)d\rvec. 
\end{equation}
Now, $e^{\mathrm{sr/lr}}_{\mathrm{eex},\sigma}(\rvec)$ can be defined as 
\begin{equation}
\begin{rcases}
	e^{\mathrm{sr/lr}}_{\mathrm{eex},\sigma}(\rvec) = -\sum_{i}^{occ} \sum_{j}^{occ} \int \frac{g^{\mathrm{sr/lr}}
(\gamma,\rvec^{\prime})\phi_{i,\sigma}(\rvec)
\phi_{j,\sigma}(\rvec)\phi_{i,\sigma}(\rvec^{\prime})\phi_{j,\sigma}(\rvec^{\prime})}{|\rvec-\rvec^{\prime}|}
d\rvec^{\prime}  \\
=-\sum_{\mu\nu}\sum_{\lambda \eta} P_{\mu\nu}^{\sigma} P_{\lambda\eta}^{\sigma} 
\int \frac{g^{\mathrm{sr/lr}}(\gamma,\rvec^{\prime})\chi_{\mu}(\rvec)
\chi_{\lambda}(\rvec)\chi_{\nu}(\rvec^{\prime})\chi_{\eta}(\rvec^{\prime})}{|\rvec-\rvec^{\prime}|}
d\rvec^{\prime},\\
g^{\mathrm{sr}}=\tilde{g}(\gamma,\rvec) \quad \mathrm{and} \quad  g^{\mathrm{lr}}=g(\gamma,\rvec),
\end{rcases}
\end{equation}
where $g^{\mathrm{sr/lr}}(\gamma,\rvec)$ is the respective RS operator for a given functional. The two expressions are 
defined in terms of (G)KS occupied MOs $\{\phi_{i,\sigma}\}$ and AOs $\{\chi_{\mu,\sigma}\}$. The complex conjugate sign 
is omitted here since the density matrix and basis are generally in real form. This definition of 
$e^{\mathrm{sr/lr}}_{\mathrm{eex},\sigma}(\rvec)$ is similar to the EEX energy density evaluation; only the 
\emph{four-center electron repulsion integrals} are modified by RS operator. 

At first glance, it seems to be computationally expensive due to the fact that it needs to be evaluated at each grid point 
with four AO indices. But, the scheme in \citep{ghosal19} shows promise in substantial 
computational cost reduction. Accordingly Eq.~(11) can be recast in the following form,
\begin{equation}
e^{\mathrm{sr/lr}}_{\mathrm{eex},\sigma}(\rvec)=-\sum_{\nu}Q_{\nu}^{\sigma}
(\rvec)M_{\nu}^{\sigma,\mathrm{sr/lr}}(\rvec).
\end{equation}
One may anticipate the construction of $e^{\mathrm{sr/lr}}_{\mathrm{eex},\sigma}(\rvec)$ in three steps of comparable 
computational cost. The first quantity $Q_{\nu}^{\sigma}(\rvec)$ may be represented as follows:
\begin{equation}
Q_{\nu}^{\sigma}(\rvec)=
\sum_{\mu} \chi_{\mu}(\rvec)P_{\mu\nu}^{\sigma},
\end{equation}
in which the density matrix is combined with AOs through a simple matrix multiplication. 
The computational scaling of this step is $\mathcal{O}(\mathrm{N_{g} N_{B}^2})$, with $\mathrm{N_g}$, 
$\mathrm{N_B}$ denoting total number of grid points and number of AO basis functions. The next crucial 
(rate-determining) step is to evaluate the SR/LR \emph{two-center electrostatic potential} (ESP) integral 
$v_{\nu\eta}^{\mathrm{sr/lr}}(\rvec)$ which is embedded in $M_{\nu}^{\sigma,\mathrm{sr/lr}}(\rvec)$ and can be defined as, 
\begin{equation}
v_{\nu\eta}^{\mathrm{sr/lr}}(\rvec)=\int\frac{g^{\mathrm{sr/lr}}(\gamma,\rvec^{\prime})\chi_{\nu}(\rvec^{\prime})
\chi_{\eta}(\rvec^{\prime})}{|\rvec-\rvec^{\prime}|}d\rvec^{\prime}.
\end{equation}
The scaling cost of this integral is $\mathcal{O}(\mathrm{N_{g} N_{B}^2})$. Usually one may perform this 
integral \emph{analytically} using primitive functions and different types of recursion algorithms, such as Obara-Saika 
\citep{obara86, obara88}, Head-Gordon-Pople \citep{gordon88} or their combination \citep{liu16}. Here, we perform FCT for 
accurate estimation of this integral, which is discussed in the following. 

The final step consists of computation of the quantity $M_{\nu}^{\sigma,\mathrm{sr/lr}}(\rvec)$ in accordance with the 
following expression: 
\begin{equation}
M_{\nu}^{\sigma,\mathrm{sr/lr}}(\rvec)=\sum_{\eta}Q_{\eta}^{\sigma}(\rvec)v_{\nu\eta}^{\mathrm{sr/lr}}.
\end{equation}
The step involves same scaling, as the ESP integral evaluation, but requires fewer steps than latter; 
needing only one multiplication and one addition at innermost loop. This also effectively provides a purely 
numerical way to compute the LR/SR exact exchange matrix, $F_{\mu \nu,\sigma}^{\mathrm{x,sr/lr}}$, which according to 
Eqs.~[11-13] can be rewritten as,
\begin{equation}
	\frac{\partial E_{\mathrm{eex},\sigma}^{\mathrm{sr/lr}}}{\partial P_{\lambda \eta}^{\sigma}}
	=F_{\mu \nu,\sigma}^{\mathrm{eex,sr/lr}}=-\int \chi_{\mu}(\rvec) M_{\nu}^{\sigma,\mathrm{sr/lr}}(\rvec) d\rvec. 
\end{equation}

Now the two-center SR/LR ESP integral, which is the core computing component, can be rewritten as, 
\begin{equation}
v^{\mathrm{sr/lr}}_{\nu \eta}(\rvec)=\int \frac{g^{\mathrm{sr/lr}}(\gamma,\rvec^{\prime}) \chi_{\nu}(\rvec^{\prime}) 
	\chi_{\eta}(\rvec^{\prime})}{|\rvec-\rvec^{\prime}|}d\rvec^{\prime}
= \int \frac{\chi_{\nu \eta}(\rvec^{\prime})g^{\mathrm{sr/lr}}(\gamma,\rvec^{\prime})}{|\rvec-\rvec^{\prime}|}
=\chi_{\nu \eta}(\rvec)\star v^{\mathrm{sr/lr}}_{\mathrm{c}}(\rvec).
\end{equation}
The last expression is in terms of convolution integral, where $\chi_{\nu \eta}$ denotes simple multiplication of two AO 
basis functions and $v^{\mathrm{sr/lr}}_{\mathrm{c}}(\rvec)$ represents the modified Coulomb interaction kernel by a given 
RS operator. Now one can invoke FCT to further simplify this integral,
\begin{equation}
v_{\nu\eta}^{\mathrm{sr/lr}}(\rvec)= \mathcal{F}^{-1}\{v_{\mathrm{c}}^{\mathrm{sr/lr}}(\mathbf{k})\chi_{\nu\eta}
	(\mathbf{k})\} \quad \mathrm{and} \quad \chi_{\nu\eta}(\mathbf{k})=\mathcal{F}\{\chi_{\nu\eta}(\rvec)\}.
\end{equation}
Here, $\mathcal{F}$ and $\mathcal{F}^{-1}$ stand for fast Fourier transformation (FFT) and inverse FFT, respectively, and 
$v_{\mathrm{c}}^{\mathrm{sr/lr}}(\mathbf{k})$ and $\chi(\mathbf{k})$ denote Fourier integrals of the modified Coulomb 
kernel and AO basis functions respectively. From the foregoing discussion, it is evident that each ESP integral involves 
only a combination of FFT (two forward and one backward transformation simultaneously) leading to 
$\mathcal{O}(\mathrm{N_{g}logN_{g}})$ scaling. Here, the computational cost remains independent of degree of 
contraction, and apart from pre-factors, $\mathrm{N_g}$, each ESP integral becomes logarithmic. This could be 
advantageous for basis sets with large degrees of contraction, for a system requiring moderate grid size. A detailed 
analysis including real-time performance of present approach with commonly used Gaussian basis functions, has already 
been enumerated in \citep{ghosal19}, and hence not repeated here.

Similarly, the SR/LR EEX energy and its contributions towards (G)KS-Fock matrix can be computed numerically in CCG as: 
\begin{equation}
\begin{rcases}
E_{\mathrm{eex},\sigma}^{\mathrm{sr/lr}}=  \frac{1}{2} h_x h_y h_z 
	\sum_g e_{\mathrm{eex},\sigma}^{\mathrm{sr/lr}}(\rvec_g), \\
	F_{\mu\nu,\sigma}^{\mathrm{\mathrm{eex},sr/lr}}
	=- h_x h_y h_z \sum_g \chi_{\mu}(\rvec_g)M_{\nu}^{\sigma,\mathrm{sr/lr}}(\rvec_g).
\end{rcases}
\end{equation}

\subsection{Optimal tuning of $\gamma$ from \emph{first principles}}
Let us consider an atom A with ground-state density $\rho_{\mathrm{A}}(\rvec)$ with 
$\gamma_{\mathrm{A}}\equiv \gamma[\rho_{\mathrm{A}}]$ and similarly for an atom B. When both the atoms are taken together 
at infinite separation, then density of the composite system will be 
$\rho_{\mathrm{A} \dots \mathrm{B}}(\rvec)=\rho_{\mathrm{A}}(\rvec)+\rho_{\mathrm{B}}(\Rvec+\rvec)$ considering $\Rvec$ is 
very large. Then the system too has a RS parameter 
$\gamma_{\mathrm{A} \dots \mathrm{B}}\equiv \gamma[\rho_{\mathrm{A} \dots \mathrm{B}}]$. That is no longer possible using 
a universal $\gamma$. Traditionally, the common RSH functionals with fixed $\gamma$ break the \emph{size-dependency} 
of total energy. Therefore, $\gamma$ should depend on system's size or more specifically be a functional of $\rho(\rvec)$, 
and it has been confirmed both from formal consideration \citep{baer05} and practical simulation for the homogeneous 
electron-gas problem \citep{livshits07}. Moreover, it has been proposed \citep{stein10} that a high level of performance 
can be achieved if one treats $\gamma$ as a system-dependent parameter tuned from \emph{first principles}. The great 
advantage of OT is that it preserves the size dependency. 

\noindent
\par 
Here we argue that it is possible to estimate $\gamma$ systematically by optimizing the total energy of a given system. 
In \citep{ghosal16, ghosal18, mandal19, ghosal19}, we have shown that within a CCG framework, for each 
combination of $N_x$, $N_y$, $N_z$, we have the self-consistent field (SCF) density and total energy of a given system 
after solving Eq.~(3). Thus, we can easily write as, 
\begin{equation}
\{N_x,N_y,N_z\}_i \iff \rho_{i}(\rvec) \iff E_{\mathrm{tot},i}, 
\end{equation}
where ``i" stands for ith combination of $N_x$, $N_y$, $N_z$ with fixed $h_r$. Further, for a given $h_r$, one can find 
out an optimal value of $\{N_x,N_y,N_z\}$ such that, 
\begin{eqnarray*}
\{N_x,N_y,N_z\}_i \equiv \{N_x,N_y,N_z\}_{\mathrm{opt}} \quad \mathrm{when} \\
\Delta E = (E_{\mathrm{tot},i}-E_{\mathrm{tot},i-1}) < \mathrm{thresh}, 
\end{eqnarray*}
where the $thresh$ is the grid accuracy for total energy convergence i.e., the energy difference between two successive 
calculations with different $N_x$, $N_y$ and $N_z$. Furthermore, each combination of $\{N_r h_r \}_{\mathrm{opt}}$, 
where $r \in \{x,y,z\}$, actually defines the optimal length of the simulation box at each direction. Among them, the 
smallest one, defined as the minimum length, where $\rho(\rvec)$ exponentially decays to zero, becomes the characteristic 
length of a given system \citep{leininger97}. 

\noindent
\par
Now for a given grid parameter, $\gamma$ can be fixed through a mathematical constraint, similar to that used in 
\citep{ghosal16, ghosal18, mandal19, ghosal19} in the context of Hartree potential and EEX contribution, 
\begin{equation}
\gamma \times L = 7, \quad L_r=N_r h_r; \quad r \in \{x,y,z\}, 
\end{equation}
where, $L$ refers to the smallest length of the simulation box. This expression is rather more 
empirical than from physical grounds. It connects the physical parameter $\gamma$, which is local in nature, 
with the numerical parameter $L$. Note that $L$ will change if the numerical grid parameter $N_r$ 
is changed, which in turn, will modify $\gamma$. Moreover, the SCF density will alter for a variation in $N_r$, which 
will again modify $\gamma$. This prompts us to write,
\begin{equation}
\{N_x,N_y,N_z\}_{\gamma} \iff \rho_{\gamma} \iff E_{\mathrm{tot},\gamma} \\
\end{equation}
Hence, each combination of $N_x$, $N_y$, $N_z$ will fix $L$, and consequently $\gamma$. Accordingly, one can find the SCF 
density and total energy of a given system for this particular $\gamma$. 

\noindent
\par
It inscribes formally the size dependency of $\gamma$ for a given system through the self-consistent density and 
consequently, with the total energy of that system. Now, an optimization of the total energy with respect to grid 
parameters gives the characteristic length of a given system, and hence $\gamma_{\mathrm{OT}}$ from Eq.~(21). Therefore, 
\begin{equation}
\gamma_{\mathrm{OT}}  \equiv \ 
\underset{N_x,N_y,N_z} {\mathrm{opt}} E_{\mathrm{tot},\gamma}, \quad \mathrm{at \ fixed} \ h_{\rvec}.
\end{equation}
Here, we have not made use of any fitting procedure for $\gamma$, with experimental results. Each system has its own 
characteristic length \citep{leininger97} and, therefore, the current procedure satisfies size dependency principle. It 
is itself sufficient to obtain the characteristic length, and hence, the optimal value of $\gamma$. This offers a simple 
machinery to estimate $\gamma_{\mathrm{OT}}$ directly in an RSH functionals from \emph{first principles}.

\noindent
\par
It is to be noted that Eq.~(23) provides a rather general mapping rule. Though, it is trivial to compute the 
characteristic length, $L$ in CCG, its generalization towards other frameworks is highly demanding. 
Through out the optimization procedure, we have kept fixed $h_r$ and the factor $7$ in Eqs.~(21) and (23). 
As pointed out in \citep{korzdorfer11}, the primary effect of $\gamma$ is to control the length 
scale for range separation, i.e., the screening of Coulomb interaction, in a particular system of interest. 
However, since $\gamma_{\text{OT}}$ has to reflect Coulomb screening, it can be expected to be sensitive to 
the size and electronic structure of the system under consideration. The formal justification of Eq.~(21) is well 
documented in \citep{martyna99} in the context of treating long-range interactions in \emph{ab initio} and 
force-field-based calculations in clusters. 
\color{red}This is based on a convergent relationship between the expressions for long-range forces in an infinitely 
replicated periodic system and those in a finite system. It basically ``screens" the interaction of the system with 
an infinite array of periodic images. They defined the screening function with a convergent parameter that controlled 
the range of interaction, and ensures that the error due to replicated periodic images in a finite system can be 
neglected. Here, we have redefined it in the desired context of exact long-range exchange of RSH functionals, based on 
the fact that we also used the same receiprocal based method to treat long-range interaction in a simulation box of 
finite length. Then, we have made an attempt to connect it through the characteristic length of a given system. 
Moreover, it has been well demonstrated that for a screening (convergence) function like $\textrm{erfc} (\gamma \rvec)$, the 
choice of $\gamma \times L \ge 7$ yields accurate results for a wide variety of systems where, $L$ denotes the smallest 
size of simulation box. At the same time, it allows efficient numerical integration on the Cartesian grid. 
\color{black}However our practical experience \citep{ghosal16, ghosal18, mandal19, ghosal19} suggests $\gamma \times L$ 
to be an optimal condition in terms of cost and accuracy. 
  
\noindent
\par 
Note that, the OT as devised in \citep{livshits07, stein10, tamblyn14, kronik12}, for the standard OT-RSH 
functionals, is generally seen as estimating $\gamma_{\mathrm{OT}}$ from \emph{first principles}. The basic difference of 
our representation and standard OT-RSH is as follows: our representation is completely different from their work requiring 
calculation only on neutral species and the optimization of total energy with respect to grid parameters. On the other hand, 
standard OT-RSH functionals used Koopmans' theorem and PWL condition on both neutral and charged species (corresponding 
cation and anion) requiring several calculations on whole range of $\gamma$ for the properties derived from orbital energies. 
We believe that our representation may be useful in the future development of \emph{ab initio} OT-RSH functionals.   

\section{Computational details}
The general framework of RSH functionals is defined in Eq.~(2), and in this rubric, we consider three well-established mode 
of partitioning. The first one, long-range correction (LC) scheme \citep{iikura01} is represented as: 
\begin{equation}
\begin{rcases}
	E_{\mathrm{xc}}^{\mathrm{LC}} = E^{\mathrm{x,sr}}_{\mathrm{dfa}}(\gamma)+E^{\mathrm{lr}}_{\mathrm{eex}}(\gamma)+ 
	E^{\mathrm{c}}_{\mathrm{dfa}}, \\
g(\gamma,\rvec)=\mathrm{erf}(\gamma\rvec) \quad \mathrm{and} \quad \tilde{g}(\gamma,\rvec)=\mathrm{erfc}(\gamma,\rvec). 
\end{rcases}
\end{equation}
The second one is that of Coulomb-attenuating method (CAM) approach \citep{yanai04} which was introduced using a more 
general form of RS operator as:
\begin{equation}
\begin{rcases}
	g_{\alpha,\beta}(\gamma,\rvec)=\alpha+\beta \ \mathrm{erf}(\gamma\rvec) \quad \mathrm{and} \quad 
	\tilde{g}_{\alpha,\beta}(\gamma,\rvec)=1-
[\alpha+\beta \ \mathrm{erf}(\gamma,\rvec)], \\
0 \leq \alpha + \beta \leq 1, \quad  0 \leq \alpha \leq 1, \quad \mathrm{and} \quad 0 \leq \beta \leq 1. 
\end{rcases}
\end{equation}
The parameter $\alpha$ allows incorporation of EEX contribution over the whole range by a factor of $\alpha$, while 
$\beta$ leads to inclusion of DFA in the entire range by a factor of $1-(\alpha+\beta)$. In the special case, 
CAM approach leads to LC with $\alpha=0, \beta=1$. These two parameters are connected to 
$a^{\mathrm{sr}}_{\mathrm{eex}}$ in a complicated manner. The last one, we consider here, is the so-called 
long-range-corrected (LRC) approach \citep{rohrdanz09}, having an extra parameter for 
$E^{\mathrm{sr}}_{\mathrm{eex}}$ as: 
\begin{equation}
\begin{rcases}
	E_{\mathrm{xc}}^{\mathrm{LRC}} = a^{\mathrm{sr}}_{\mathrm{eex}}E^{\mathrm{sr}}_{\mathrm{eex}}(\gamma)+
	(1-a^{\mathrm{sr}}_{\mathrm{eex}})E^{\mathrm{x,sr}}_{\mathrm{dfa}}(\gamma)
	+E^{\mathrm{lr}}_{\mathrm{eex}}(\gamma)+E^{\mathrm{c}}_{\mathrm{dfa}}, \\
g(\gamma,\rvec)=\mathrm{erf}(\gamma\rvec) \quad \mathrm{and} \quad \tilde{g}(\gamma,\rvec)=\mathrm{erfc}(\gamma,\rvec). 
\end{rcases}
\end{equation}
This parameter $a^{\mathrm{sr}}_{\mathrm{eex}}$ allows to incorporate a desired amount of $E^{\mathrm{sr}}_{\mathrm{eex}}$ 
by a factor of $a^{\mathrm{sr}}_{\mathrm{eex}}$. In special case, the LRC approach with $a^{\mathrm{sr}}_{\mathrm{eex}}=0$ 
leads to LC. A variety of other partitioning schemes including different RS operators have also been 
explored in the literature \citep{chai08a, chai08b, peverati12, vikramaditya18, chan19}, mainly in connection with 
thermochemistry and reaction barrier heights.

An important aspect of RSH functionals is the successful development of $E^{\mathrm{x,sr}}_{\mathrm{dfa}}$. Several schemes 
have been proposed in the past, such as, based on model exchange hole \citep{iikura01, henderson08}, adiabatic connection 
theorem \citep{baer05, cohen07}, exchange energy density \citep{chai08b,lin12}. Here, we have used the formulation of 
\citep{iikura01}, applicable to any LDA or GGA type DFAs, which involves modified Fermi wave vector in exchange enhancement 
factor. This was later adopted to develop CAM-B3LYP functional \citep{yanai04} using a more general form of RS operator as 
defined in Eq.~(25). Accordingly, the SR GGA-exchange energy can be cast as:
\begin{equation}
\begin{rcases}
E^{\mathrm{x,sr}}_{\mathrm{gga}}=-\frac{1}{2}\sum_{\sigma}\int\rho_{\sigma}^{\frac{4}{3}}
	K_{\mathrm{gga},\sigma}^{\mathrm{x,sr}}d\rvec, \\ 
K_{\mathrm{gga},\sigma}^{\mathrm{x,sr}}=K_{\mathrm{gga},\sigma}^{\mathrm{x}}\bigg{[}(1-\alpha)
-\beta \bigg\{\frac{8}{3}a_{\sigma} \big{[} \sqrt{\pi} \ \mathrm{erf} \big{(}\frac{1}{2a_{\sigma}}
	+2a_{\sigma}(b_{\sigma}-c_{\sigma})
\big{)} \big{]} \bigg \}\bigg{]}, \\
a_{\sigma}=\frac{\gamma}{2K_{\mathrm{gga},\sigma}^{\mathrm{f}}}, b_{\sigma}=
	\mathrm{exp}\bigg{(}-\frac{1}{4a_{\sigma}^{2}}\bigg{)}, 
c_{\sigma}=2a_{\sigma}^{2}b_{\sigma}+\frac{1}{2},K_{\mathrm{gga},\sigma}^{\mathrm{f}}
	=\bigg{(}\frac{9\pi}{K_{\mathrm{gga},\sigma}^{\mathrm{x}}}
\bigg{)}^{\frac{1}{2}}\rho_{\sigma}^{\frac{1}{3}}, 
\end{rcases}
\end{equation}
where, $K_{\mathrm{gga},\sigma}^{\mathrm{x}}$ is the usual enhancement factor. The average relative momentum for GGA, 
$K_{\mathrm{gga},\sigma}^{\mathrm{f}}$ is used to define the modified GGA-enhancement factor, 
$K_{\mathrm{gga},\sigma}^{\mathrm{x,sr}}$. It is easily seen that Eq.~(27) reproduces the original GGA DFAs for 
$\gamma=\alpha=0$. The corresponding potential is evaluated using the modified GGA-enhancement factor as it was done 
for standard GGA DFAs \citep{johnson93}. Further development of more balanced SR DFAs can be found in 
\citep{iikura01, henderson08, baer05, cohen07, chai08b, lin12}. Note that, the original LRC scheme was proposed based 
on Perdew-Burke-Ernzerhof (PBE) exchange hole, which satisfies the above constraints \citep{ernzerhof98} at all $\gamma$.
 
The three distinct kind of RSH functionals (LC, CAM and LRC) as mentioned above are used in our calculations keeping 
the mode of partitioning and RS operator fixed as in the original articles. As mentioned earlier, the default values of 
$\gamma$, and other auxiliary parameters ($a^{\mathrm{sr}}_{\mathrm{eex}}, \alpha, \beta$) were obtained semi-empirically 
by fitting the reference data. Here, however, we follow the strategy of Sec.~II to determine $\gamma_{\mathrm{OT}}$. 
These are implemented in case of five representative set of functionals containing a variable amount of SR/LR exact 
exchange with SR DFA exchange and conventional correlation functional in our in-house pseudopotential (G)KS-DFT program 
in CCG, InDFT \citep{indft19}. We consider the LC-BLYP \citep{tawada04} and LC-PBE \citep{iikura01,perdew96} functionals 
from LC-hybrid group with $\gamma=0.33$ and $\gamma=0.30$, respectively. Moreover, CAM-B3LYP \citep{yanai04} employing 
$\alpha=0.19, \beta=0.46$, $\gamma=0.33$ and CAM-PBE0 \citep{lange08} with $\alpha=0.25, \beta=0.75$, $\gamma=0.30$ are 
used for CAM-hybrid group. The original LRC-$\omega$PBEh functional \citep{rohrdanz09} with 
$a^{\mathrm{x,sr}}=0.2$, $\gamma=0.2$ is used for LRC-hybrid group with slight modification. Here, it is denoted as 
LRC-$\omega$PBEh$^{\star}$. It is superscripted with a $\star$ to differentiate from the original. The only difference is about 
the construction of SR DFA-exchange. In \citep{rohrdanz09}, it is based on PBE exchange hole. 
In present work, we have used the procedure from one-particle density matrix, as we done for LC and CAM categories, to make 
them consistent with each other. All the parameters except $\gamma$ are kept fixed as in the original article, 
and these functionals are denoted with the subscript ``ot". Two global hybrid functionals, namely, B3LYP, and PBE0, 
containing a variable amount of EEX energy with a conventional DFA are also considered side by side \citep{stephens94, perdew96a}. 
 
\begingroup                      
\squeezetable
\begin{table}      
\caption{\label{tab:table1} Convergence of CAM-B3LYP$^{\dagger}$ energy of HCl in the grid 
($h_r=0.3$). Results are in a.u.}
\begin{ruledtabular}
\begin{tabular} {ccccccccc}
 \multicolumn{4}{c}{Set~I }  & & \multicolumn{4}{c}{Set~II}      \\
\cline{1-4} \cline{6-9} 
$N_x$   &   $N_y$   &    $N_z$  &  $ \langle E \rangle $ &    & $N_x$   &   $N_y$   &    $N_z$  &  $ \langle E \rangle$ 
\\
\cline{1-9}
32 & 32 & 32 & $-$15.48226 &  &32 & 36 & 60 & $-$15.50095 \\
-- & -- & 36 & $-$15.49414 &  &32 & 44 &  -- & $-$15.50175 \\
-- & -- & 40 & $-$15.49695 &  &32 & 48 &  -- & $-$15.50175 \\
-- & -- & 44 & $-$15.49760 &  &32 & 52 &  -- & $-$15.50178 \\
-- & -- & 48 & $-$15.49775 &  &32 & -- &  -- & $-$15.50178 \\
-- & -- & 52 & $-$15.49777 &  &36 & -- & -- & $-$15.50495 \\
-- & -- & 56 & $-$15.49778 &  &44 & -- & -- & $-$15.50575 \\
-- & -- & 60 & $-$15.49778 &  &48 & -- & -- & $-$15.50578 \\
-- & -- & 64 & $-$15.49778 &  &52 & -- & -- & $-$15.50578 \\
\end{tabular}
\end{ruledtabular}
\begin{tabbing}
$^{\dagger}$Energy from GAMESS package \cite{schmidt93}: $-$15.50592 a.u. \\
\end{tabbing}
\end{table}
\endgroup

We employ following effective core potential (ECP) basis sets: SBKJC \citep{stevens84} for species containing Group-II 
elements and LANL2DZ \citep{hay85c} for Group-III or higher group elements. These are adopted from EMSL Basis Set Library 
\citep{feller96}. All one-electron integrals are generated by standard recursion relations \citep{obara86} using Cartesian 
Gaussian-type orbitals as primitive basis functions. The norm-conserving pseudopotential matrix elements in contracted 
basis are imported from GAMESS \citep{schmidt93} suite of program package. The relevant LDA- and GGA-type of functionals 
in connection with B3LYP, and PBE0 are: (i) Vosko-Wilk-Nusair (VWN)--with the homogeneous electron gas correlation 
proposed in parametrization formula V \citep{vosko80} (ii) B88--incorporating Becke \citep{becke88a} semi-local exchange 
(iii) Lee-Yang-Parr (LYP) \citep{lee88} semi-local correlation (iv) PBE \citep{perdew96} for semi-local exchange and 
correlation. The modified SR-exchange B88 from Eq.~(27) and LYP correlation are invoked for LC-BLYP and CAM-B3LYP, whereas 
the modified SR-exchange PBE from Eq.~(27) and PBE correlation for LC-PBE, CAM-PBE0, LRC-$\omega$PBEh$^{\star}$. All 
correlation functionals are directly adopted from density functional repository program \citep{repository}. The convergence 
criteria imposed in this communication are slightly tighter than our earlier work \citep{roy08,roy08a,roy10,roy11}; this is 
to generate a more accurate orbital energy. Changes in following quantities were followed during the SCF process, 
\emph{viz.,} (i) orbital energy difference between two successive iterations and (ii) absolute deviation in a density matrix 
element. They both are required to remain below a prescribed threshold set to $10^{-8}$ a.u.; this ensured that total energy 
maintained a convergence of at least this much, in general. In order to perform discrete Fourier transform, standard FFTW3 
package \citep{fftw05} is invoked. The resulting generalized matrix-eigenvalue problem is solved through standard LAPACK 
routine \citep{anderson99} accurately and efficiently. All molecular calculations are performed in their experimental 
geometries, taken from NIST database \citep{johnson16}. Other details including scaling properties may be found in references 
\citep{roy08, roy08a, roy10, roy11, ghosal16, ghosal18, mandal19, ghosal19, ghosal21}.

\begingroup                      
\squeezetable
\begin{table}     
\caption{\label{tab:table2} Convergence of LC-BLYP$^{\dagger}$ energy of H$_2$S in the grid ($h_r=0.3$). 
All results are in a.u.}
\begin{ruledtabular}
\begin{tabular} {ccccccccc}
 \multicolumn{4}{c}{Set~I }  & & \multicolumn{4}{c}{Set~II}      \\
\cline{1-4} \cline{6-9} 
$N_x$   &   $N_y$   &    $N_z$  &  $ \langle E \rangle $ &    & $N_x$   &   $N_y$   &    $N_z$  &  $ \langle E \rangle$ 
\\
\cline{1-9}
32 & 32 & 32 & $-$11.14114 &  &32 & 40 & 60 & $-$11.17356 \\
-- & -- & 36 & $-$11.15210 &  &32 & 44 &  -- & $-$11.17456 \\
-- & -- & 40 & $-$11.15508 &  &32 & 52 &  --  & $-$11.17486 \\
-- & -- & 44 & $-$11.15583 &  &32 & 56 &  --  & $-$11.17487 \\
-- & -- & 48 & $-$11.15600 &  &32 & 60 &  -- & $-$11.17487 \\
-- & -- & 52 & $-$11.15604 &  &36 & -- &  -- & $-$11.18494 \\
-- & -- & 56 & $-$11.15605 &  &48 & -- &  -- & $-$11.18850 \\
-- & -- & 60 & $-$11.15606 &  &56 & -- & -- & $-$11.18855 \\
-- & -- & 64 & $-$11.15606 &  &60 & -- &  -- & $-$11.18855 \\
\end{tabular}
\end{ruledtabular}
\begin{tabbing}
$^{\dagger}$Energy from GAMESS package \cite{schmidt93}: $-$11.18865 a.u. \\
\end{tabbing}
\end{table}
\endgroup

\section{Results and discussion}
Before proceeding for main results, it may be appropriate first to discuss the grid optimization on the convergence of 
total energy, $E_{\mathrm{tot}}$, which plays a vital role for practical implementation of Eq.~(23). For illustration, 
we choose two RSH functionals, namely, CAM-B3LYP and LC-BLYP. All the calculations presented in this section are performed 
through our in-house pseudopotential (G)KS-DFT program in CCG \citep{indft19}. At first, Table~I shows this for CAM-B3LYP 
for HCl at its experimental geometry (bond length $1.2746$\textup{\AA}). The total energy is provided in non-uniform grid 
with respect to \emph{sparsity} (regulated by $N_x, N_{y}, N_{z}$) for a fixed grid spacing (determined by $h_r$, chosen as 
0.3), employing fixed parameter $\alpha=0.19$, $\beta=0.46$ and $\gamma=0.33$ from Eq.~(25). Following our previous works 
\citep{ghosal16, ghosal18, ghosal19} we first vary $N_z$, the number of grid points along inter-nuclear axis, keeping the 
same along $xy$ plane static at certain reasonable value, say $N_x=N_y=32$. As $N_z$ is gradually increased from 32 to 64 
with an increment of 4, there is a smooth convergence in energy at around $N_z=60$, with a difference (we term it as grid 
accuracy) of about $ 5 \times 10^{-6}$ a.u., between two successive steps. In the beginning, when $N_z$ goes through 
$40-44-48-52$, one notices slow improvement in energy; after that it eventually attains the convergence for $N_z$ at around 
$60$ for a particular value of grid accuracy. Then in Set~II in right-hand side, we vary sequentially $N_{y}$ and $N_{x}$ 
along $xy$ plane keeping first $N_z$ and $N_x$ fixed at its previously determined value of $60$ and $32$ respectively, and 
then vary $N_x$ keeping $N_z$ and $N_y$ fixed at $60$ and $56$ respectively. Now we can see that the convergence in energy 
takes place at $N_{x}=52$ and $N_{y}=56$ with same grid accuracy of Set~I. For sake of completeness, the respective energy 
is also quoted from GAMESS \cite{schmidt93} in footnote. Next we move toward the implementation of LC-BLYP using the same 
procedure mentioned above. For this, we consider a non-linear triatomic molecule H$_2$S as a specimen case. The stability of 
our current implementation through the total energy convergence is illustrated in Table II at a grid spacing of $h_r= 0.3$ 
using the same convergence criteria as imposed in Table~I. Here also the respective energy value is quoted from GAMESS in 
footnote. The performance of our calculated energies from Tables~I and II are quite accurate with the reference results for 
a given grid accuracy. 

\begingroup                 
\squeezetable
\begin{table}     
\caption{\label{tab:table3} Ionization energies, $-\epsilon_{\mathrm{HOMO}}$ for selected atoms in eV.}
\resizebox{1.0\textwidth}{!}{
\begin{tabular} {l|ccccc|ccccccc|c}
\cline{1-14}
Atom & B3LYP & LC-BLYP & LC-BLYP$_{\mathrm{ot}}$ & CAM-B3LYP & CAM-B3LYP$_{\mathrm{ot}}$ & PBE$0$ & LC-PBE 
& LC-PBE$_{\mathrm{ot}}$ & CAM-PBE$0$ & CAM-PBE$0$$_{\mathrm{ot}}$ & LRC-$\omega$PBEh$^{\star}$ & 
LRC-$\omega$PBEh$^{\star}$$_{\mathrm{ot}}$ 
& Expt.$^{\dagger}$ \\ 
\cline{1-14}
Be & 6.23 & 8.52 & 8.50 & 7.64 & 7.63 & 6.50 & 8.58 & 8.67 & 8.71 & 8.78 & 8.23 & 8.75 & 9.32 \\
B  & 4.98 & 7.55 & 7.85 & 6.63 & 6.77 & 5.25 & 7.45 & 7.93 & 7.79 & 8.16 & 7.06 & 8.11 & 8.30 \\
C  & 7.12 & 9.88 & 10.67 & 8.96 & 9.32 & 7.54 & 9.79 & 10.65 & 10.39 & 11.04 & 9.43 & 10.96 & 11.26 \\
N  & 9.52 & 12.27 & 13.53 & 11.47 & 12.05 & 10.09 & 12.22 & 13.79 & 13.12 & 14.30 & 11.98 & 14.20 & 14.53 \\
O  & 8.83 & 11.42 & 12.97 & 10.77 & 11.49 & 9.19 & 11.13 & 13.00 & 12.10 & 13.50 & 10.95 & 13.40 & 13.62 \\
Al & 3.51 & 5.59 & 5.63 & 4.79 & 4.81 & 3.83 & 5.74 & 5.87 & 5.87 & 5.96 & 5.46 & 6.09 & 5.99 \\
Si & 5.27 & 7.67 & 7.72 & 6.78 & 6.80 & 5.68 & 7.81 & 8.01 & 8.06 & 8.22 & 7.46 & 8.18 & 8.15 \\
P  & 6.93 & 9.46 & 9.77 & 8.57 & 8.71 & 7.43 & 9.63 & 10.13 & 10.03 & 10.40 & 9.29 & 10.35 & 10.49 \\
S  & 6.82 & 9.37 & 9.82 & 8.49 & 8.69 & 7.14 & 9.34 & 10.00 & 9.75 & 10.24 & 8.97 & 10.19 & 10.36 \\
Cl & 8.96 & 11.63 & 12.37 & 10.75 & 11.10 & 9.36 & 11.58 & 12.57 & 12.17 & 12.92 & 11.25 & 12.85 & 12.97 \\
Ga & 3.44 & 5.49 & 5.53 & 4.71 & 4.72 & 3.76 & 5.65 & 5.78 & 5.77 & 5.87 & 5.37 & 5.85 & 6.00 \\
Ge & 4.97 & 7.27 & 7.32 & 6.41 & 6.43 & 5.37 & 7.44 & 7.62 & 7.66 & 7.79 & 7.12 & 7.76 & 7.90 \\
As & 6.60 & 9.05 & 9.22 & 8.17 & 8.24 & 7.09 & 9.24 & 9.57 & 9.57 & 9.82 & 8.91 & 9.77 & 9.82 \\
Se & 6.36 & 8.83 & 9.10 & 7.94 & 8.06 & 6.67 & 8.84 & 9.28 & 9.15 & 9.49 & 8.47 & 9.47 & 9.75 \\
Br & 8.19 & 10.78 & 11.10 & 9.87 & 10.02 & 8.56 & 10.77 & 11.30 & 11.20 & 11.60 & 10.41 & 11.54 & 11.81 \\
\cline{1-14}
MAE & 3.50 & 1.03 & 0.61 & 1.89 & 1.69 & 3.12 & 1.00 & 0.41 & 0.60 & 0.15 & 1.33 & 0.20 &  \\
ME & 3.50 & 1.03 & 0.61 & 1.89 & 1.69 & 3.12 & 1.00 & 0.41 & 0.60 & 0.15 & 1.33 & 0.19 &  \\
$\Upsilon$ &  &  & 1.69 &  & 1.12 &  &  & 2.44 &  & 4.00 &  & 6.65 & \\
\cline{1-14}
\end{tabular}}
\begin{tabbing}
$^{\dagger}$Optical spectroscopy \citep{johnson16}. 
\= \hspace{160pt}
$^{\ddagger}$$\Upsilon$: Ratio between MAE value of RSH and OT-RSH. 
\end{tabbing}
\end{table}
\endgroup

To broaden the scope of applicability, we now focus on certain properties derived from frontier orbital energies. 
Henceforth we use the same grid optimization procedure through Eq.~(23) maintaining the same grid accuracy of Tables~I 
and II. To put things in perspective, we categorize the five functionals into two distinct blocks (B3LYP and PBE0) such 
that those RSH functionals containing ``B88" exchange and ``LYP" correlation belong to B3LYP and those including ``PBE" 
exchange and correlation belong to PBE0 block. Moreover, the mean absolute error (MAE) and mean error (ME) from 
statistical analysis have been provided to facilitate a detailed comparison with the available \emph{ab initio}/experimental 
results. 

\subsection{Ionization energies}
The physical interpretation of KS frontier orbital and its energies as single-particle quantities is still far from 
straightforward, even if we know the exact XC potential, except the highest occupied molecular orbital (HOMO). It can be 
assigned using the ``KS analogue of Koopmans' theorem in Hartree-Fock theory" \citep{perdew82, levy84, perdew97} and 
accordingly, one can write as,
\begin{equation}
\mathrm{IE}(M)=-\epsilon_{\mathrm{HOMO}},
\end{equation}
where, IE($M$) be the first ionization energy of a given $M$-electron system. In the context of LDA or GGA-type DFAs, Eq.~(28)
is no longer be a valid statement; the HOMO energy is usually underestimated. Moreover, this will not work for other functionals 
outside of KS regime, particularly in which we are interested in this communication. The RSH functionals, \emph{in principle}, 
have correct asymptotic behavior at LR region, but the essence of HOMO and its energy is possible only through (G)KS version of 
Koopmans' theorem. It has been proved that for the specific case of an EEX operator, it is still possible to identify the 
(G)KS HOMO energy with $-\mathrm{IE}(M)$ \citep{gorling97}, and accordingly 
\begin{equation}
	\mathrm{IE}(M)=-\epsilon_{\mathrm{HOMO}}^{\gamma}.
\end{equation}. 
Like KS mapping, the (G)KS map is 
not unique, and considering RSH functional, any choice of $\gamma$ generates a legitimate approximate (G)KS map. The obvious 
question is then, whether RSH functionals with fixed values of $\gamma$ can approximate (G)KS HOMO energy accurately with 
$-\mathrm{IE}(M)$, irrespective of systems of interest? Therefore, the comparison of the (G)KS HOMO energy with experimental 
$-\mathrm{IE}(M)$ is a good test in determining $\gamma_{\mathrm{OT}}$ through Eq.~(23).

\begingroup                 
\squeezetable
\begin{table}     
\caption{\label{tab:table4} Ionization energies, $-\epsilon_{\mathrm{HOMO}}$ for selected molecules in eV.}
\resizebox{1.0\textwidth}{!}{
\begin{tabular} {l|ccccc|ccccccc|c}
\cline{1-14}
System & B3LYP & LC-BLYP & LC-BLYP$_{\mathrm{ot}}$ & CAM-B3LYP & CAM-B3LYP$_{\mathrm{ot}}$ & PBE$0$ & LC-PBE 
& LC-PBE$_{\mathrm{ot}}$ & CAM-PBE$0$ & CAM-PBE$0$$_{\mathrm{ot}}$ & LRC-$\omega$PBEh$^{\star}$ & 
LRC-$\omega$PBEh$^{\star}$$_{\mathrm{ot}}$ 
& Expt.$^{\dagger}$ \\ 
\cline{1-14}
N$_2$ & 11.48 & 14.25 & 14.81 & 13.45 & 13.71 & 11.83 & 13.94 & 14.82 & 14.88 & 15.54 & 13.71 & 15.40 & 15.60 \\
Si$_2$ & 5.71 & 7.91 & 7.90 & 7.09 & 7.08 & 6.03 & 8.10 & 8.16 & 8.21 & 8.26 & 7.82 & 8.24 & 7.90 \\
P$_2$ & 7.89 & 10.33 & 10.50 & 9.46 & 9.54 & 8.23 & 10.33 & 10.69 & 10.70 & 10.97 & 10.01 & 10.92 & 10.62 \\
S$_2$ & 6.66 & 8.49 & 8.69 & 7.59 & 7.69 & 7.06 & 8.45 & 8.86 & 8.91 & 9.21 & 8.11 & 9.14 & 9.40 \\
Cl$_2$ & 8.55 & 11.21 & 11.78 & 10.34 & 10.60 & 8.98 & 11.18 & 11.99 & 11.79 & 12.40 & 10.87 & 12.32 & 11.48 \\

CO & 10.30 & 13.10 & 13.58 & 12.21 & 12.44 & 10.60 & 12.82 & 13.59 & 13.56 & 14.14 & 12.49 & 14.03 & 14.10 \\ 
NaCl & 5.79 & 8.37 & 7.67 & 7.44 & 7.12 & 6.12 & 8.33 & 7.81 & 8.79 & 8.45 & 8.26 & 8.33 & 9.80 \\ 
HCl & 8.91 & 11.60 & 12.35 & 10.71 & 10.98 & 9.32 & 11.55 & 12.54 & 12.15 & 12.90 & 11.22 & 12.83 & 12.79 \\ 

CO$_2$ & 10.30 & 13.10 & 13.58 & 12.21 & 12.44 & 10.61 & 12.82 & 13.59 & 13.56 & 14.14 & 12.49 & 14.03 & 14.10 \\
H$_2$O & 8.26 & 10.97 & 11.51 & 10.21 & 10.45 & 8.63 & 10.71 & 11.55 & 11.65 & 12.28 & 10.51 & 12.14 & 12.62 \\
H$_2$S & 7.12 & 9.77 & 10.11 & 8.84 & 8.99 & 7.48 & 9.74 & 10.29 & 10.18 & 10.60 & 9.37 & 10.54 & 10.48 \\
C$_2$H$_2$ & 7.94 & 10.72 & 11.04 & 9.68 & 9.83 & 8.33 & 10.71 & 11.24 & 11.08 & 11.47 & 10.28 & 11.42 & 11.40 \\
NH$_3$ & 6.65 & 9.39 & 9.86 & 8.53 & 8.75 & 7.02 & 9.21 & 9.96 & 9.95 & 10.51 & 8.92 & 10.40 & 10.82 \\
PH$_3$ & 7.25 & 9.77 & 9.96 & 8.88 & 8.97 & 7.55 & 9.72 & 10.10 & 10.11 & 10.40 & 9.37 & 10.34 & 9.89 \\

CH$_4$ & 10.48 & 13.19 & 13.78 & 12.30 & 12.57 & 10.85 & 13.06 & 13.91 & 13.68 & 14.32 & 12.73 & 14.24 & 13.6 \\
CH$_3$Cl & 8.02 & 10.67 & 11.04 & 9.76 & 9.94 & 8.39 & 10.60 & 11.20 & 11.14 & 11.59 & 10.27 & 11.51 & 11.29 \\
SiH$_4$ & 9.71 & 12.28 & 12.37 & 11.40 & 11.44 & 10.02 & 12.19 & 12.49 & 12.65 & 12.88 & 11.84 & 12.80 & 12.30 \\
C$_2$H$_4$ & 7.27 & 10.00 & 10.18 & 8.95 & 9.03 & 7.67 & 10.06 & 10.41 & 10.34 & 10.60 & 9.62 & 10.56 & 10.51 \\
C$_2$H$_6$ & 9.12 & 11.71 & 11.95 & 10.86 & 10.97 & 9.46 & 11.59 & 12.06 & 12.16 & 12.52 & 11.28 & 12.43 & 11.99 \\
Si$_2$H$_6$ & 8.23 & 10.62 & 10.59 & 9.76 & 9.74 & 8.54 & 10.62 & 10.75 & 10.94 & 11.04 & 10.30 & 10.98 & 10.53 \\
\cline{1-14}
MAE & 3.28 & 0.70 & 0.47 & 1.58 & 1.45 & 2.93 & 0.80 & 0.44 & 0.43 & 0.38 & 1.09 & 0.37 &  \\
ME & 3.28 & 0.69 & 0.40 & 1.58 & 1.45 & 2.93 & 0.77 & 0.26 & 0.24 & -0.15 & 1.09 & -0.07 &  \\
$\Upsilon$ &  &  & 1.49 &  & 1.09 &  &  & 1.8 &  & 1.13 &  & 2.95 & \\
\cline{1-14}
\end{tabular}}
\begin{tabbing}
$^{\dagger}$Photo-electron spectroscopy \citep{johnson16}. 
\= \hspace{140pt}
$^{\ddagger}$$\Upsilon$: Ratio between MAE value of RSH and OT-RSH. 
\end{tabbing}
\end{table}
\endgroup

To evaluate the performance of our approach on (G)KS HOMO energy, we consider 15 atoms and 20 molecules in our dataset; 
such molecules are taken from G1 database \citep{pople89}. The calculated negative (G)KS HOMO energies for 12 functionals 
(considering B3LYP and PBE0 blocks) including experimental $\mathrm{IE}(M)$ from NIST database \citep{johnson16} are 
collected in Tables~III and IV. A glance at both MAE and ME of Table~III, reveals that these are more or less close to 
each other for all functionals and also having a positive ME. These clearly indicate an underestimation of IE from the 
experimental values. But, at the same time, these also provide a scope to further reduce the error systematically within our 
proposed scheme, which is discussed later. Now, the fruitfulness of OT-RSH functionals is reflected through $\Upsilon$, a 
ratio of MAE between RSH and its respective OT-RSH counterpart. The value of $\Upsilon$ for all five functionals are quoted 
in Table~III. Accordingly, LRC-$\omega$PBEh$^{\star}$$_{\mathrm{ot}}$ has the largest value of $6.65$ i.e., it reduces the 
error $6.65$ times more relative to that of LRC-$\omega$PBEh$^{\star}$. Now, based on this, one can arrange the five 
functionals in descending order of performance as: LRC-$\omega$PBEh$^{\star}$$_{\mathrm{ot}}$, CAM-PBE$0_{\mathrm{ot}}$, 
LC-PBE$_{\mathrm{ot}}$, LC-BLYP$_{\mathrm{ot}}$, CAM-B3LYP$_{\mathrm{ot}}$. But, if we compare only MAE, then the 
performances of LRC-$\omega$PBEh$^{\star}$$_{\mathrm{ot}}$ and CAM-PBE$0_{\mathrm{ot}}$ are very close to each other, having 
MAE of $0.20$ and $0.15$. Again, MAE of OT-RSH from PBE0 block is less than $0.5$ eV, being more accurate 
than OT-RSH functionals from B3LYP block. 

For molecules, a closer look at both MAE and ME in Table~IV, reveals that these are also more or less equal to each other 
except for CAM-PBE$0_{\mathrm{ot}}$ and LRC-$\omega$PBEh$^{\star}$$_{\mathrm{ot}}$. The MAE and ME for latter two functionals 
are not close to each other and also these have negative ME. The systematic underestimation of IE of atoms does not happen 
in most of the cases here, so nicely. Here also, the comparison of respective $\Upsilon$ is quite useful. Accordingly, 
LRC-$\omega$PBEh$^{\star}$$_{\mathrm{ot}}$ has largest value of $2.95$ i.e., it reduces the error $2.95$ times more relative 
to that of LRC-$\omega$PBEh$^{\star}$. In this occasion, the functionals can be arranged in the following descending order of 
performance as: LRC-$\omega$PBEh$^{\star}$$_{\mathrm{ot}}$, LC-PBE$_{\mathrm{ot}}$, CAM-PBE$0_{\mathrm{ot}}$, 
LC-BLYP$_{\mathrm{ot}}$, CAM-B3LYP$_{\mathrm{ot}}$. The relative performance of LC-PBE$_{\mathrm{ot}}$ and 
LC-BLYP$_{\mathrm{ot}}$ are much better than CAM-PBE0$_{\mathrm{ot}}$ and CAM-B3LYP$_{\mathrm{ot}}$ respectively. It may be 
due to fact that the auxiliary parameters ($\alpha, \beta, a^{\mathrm{x,sr}}$) may have some sensitivity during 
self-consistent tuning process, which have been kept fixed. However, the performance of CAM-PBE0 is 
evidently better than CAM-B3LYP. This may be possibly due to the fact that, it is necessary to have $\alpha + \beta=1$ for 
asymptotically correct 
Coulomb-attenuating method, at least for atoms. But the compatibility of $\gamma_{\text{OT}}$ with these auxiliary 
parameters present in RSH functionals, should be given due consideration during the optimization procedure, and it is indeed 
important for molecular systems. In fact, caution should be exercised during their implementation, and an elaborate analysis is 
required to uncover the compatibility of these parameters with $\gamma_{\mathrm{OT}}$. Nevertheless, the accuracy 
of OT-RSH functionals is always improved than that of RSH functionals for all the species and blocks of interest. 

\begingroup                 
\squeezetable
\begin{table}     
\caption{\label{tab:table5} The relative error in $\epsilon_{\mathrm{LUMO}}$ as compared to $\Delta \mathrm{SCF}$. 
All results are in eV.}
\resizebox{1.0\textwidth}{!}{
\begin{tabular} {l|ccccc|ccccccc}
\cline{1-13}
Atom & B3LYP & LC-BLYP & LC-BLYP$_{\mathrm{ot}}$ & CAM-B3LYP & CAM-B3LYP$_{\mathrm{ot}}$ & PBE$0$ & LC-PBE 
& LC-PBE$_{\mathrm{ot}}$ & CAM-PBE$0$ & CAM-PBE$0$$_{\mathrm{ot}}$ & LRC-$\omega$PBEh$^{\star}$ & 
LRC-$\omega$PBEh$^{\star}$$_{\mathrm{ot}}$ 
 \\ 
\cline{1-13}
B  & 2.69 & 0.39 & 0.26 & 1.28 & 1.20 & 2.46 & 0.42 & 0.15 & 0.16 & 0.03 & 0.72 & 0.03  \\
C  & 3.33 & 0.85 & 0.39 & 1.73 & 1.52 & 3.07 & 1.03 & 0.41 & 0.49 & 0.05 & 1.28 & 0.12 \\
O  & 4.13 & 1.65 & 0.51 & 2.37 & 1.86 & 3.59 & 1.69 & 0.26 & 0.78 & 0.25 & 1.79 & 0.15  \\
Al & 2.14 & 0.19 & 0.23 & 1.00 & 1.00 & 1.95 & 0.15 & 0.08 & 0.06 & 0.02 & 0.37 & 0.03  \\
Si & 2.71  & 0.41 & 0.37 & 1.33 & 1.33 & 2.49 & 0.47 & 0.31 & 0.28 & 0.15 & 0.78 & 0.19 \\
P  & 3.00 & 0.60 & 0.41 & 1.56 & 1.49 & 2.58 & 0.46 & 0.12 & 0.23 & 0.01 & 0.79 & 0.01  \\
S  & 3.30 & 0.76 & 0.49 & 1.74 & 1.60 & 2.91 & 0.75 & 0.29 & 0.40 & 0.06 & 1.08 & 0.10  \\
Cl & 3.81 & 1.19 & 0.68 & 2.14 & 1.90 & 3.43 & 1.28 & 0.55 & 0.77 & 0.23 & 1.59 & 0.30  \\
Ga & 2.15 & 0.21 & 0.21 & 1.01 & 0.98 & 1.95 & 0.16 & 0.09 & 0.07 & 0.02 & 0.38 & 0.03  \\
Ge & 2.57 & 0.34 & 0.34 & 1.24 & 1.24 & 2.36 & 0.38 & 0.27 & 0.21 & 0.13 & 0.67 & 0.16  \\
As & 2.79 & 0.49 & 0.39 & 1.42 & 1.37 & 2.38 & 0.32 & 0.10 & 0.15 & 0.01 & 0.63 & 0.01  \\
Se & 3.05 & 0.61 & 0.45 & 1.57 & 1.50 & 2.69 & 0.56 & 0.26 & 0.30 & 0.07 & 0.90 & 0.11  \\
Br & 3.34 & 0.79 & 0.51 & 1.77 & 1.64 & 3.00 & 0.84 & 0.38 & 0.48 & 0.13 & 1.18 & 0.18  \\
\cline{1-13}
MAE & 3.00 & 0.65 & 0.40 & 1.55 & 1.43 & 2.68 & 0.65 & 0.25 & 0.34 & 0.09 & 0.94 & 0.11 \\
\cline{1-13}
\end{tabular}}
\end{table}
\endgroup
 
\subsection{Fundamental Gaps}
The fundamental gap (FG), $E_{\mathrm{FG}}$ is defined by ``charged excitation" as, 
\begin{equation}
E_{\mathrm{FG}} = \mathrm{IE}(M)-\mathrm{EA}(M),
\end{equation}
with EA($M$) being the first electron affinity of an $M$-electron system. According to quasi-particle theory \citep{landau57}, 
the maximum energy required for simultaneous creation of non-interacting quasi-particle and quasi-hole will be given by FG 
\citep{kronik12}. In principle, it is possible to compute FG from the energy difference between HOMO and lowest unoccupied 
molecular orbital (LUMO) of KS-DFT calculation. Even if exact XC potential were known, KS-HOMO will correspond to the 
lowest quasi-hole excitation energy, but KS-LUMO will not reflect the lowest quasi-electron excitation energy, due to presence 
of derivative discontinuity (DD) in XC potential at an integer particle number. Therefore, it can be redefined as, 
\begin{equation}
	E_{\mathrm{FG}} = \mathrm{IE}(M)-\mathrm{EA}(M)=\epsilon_{\mathrm{LUMO}}-\epsilon_{\mathrm{HOMO}}+\Delta_{\mathrm{xc}},
\end{equation}
where $\Delta_{xc}$ is denoted as DD. Therefore, the KS-DFT framework is inherently incompatible with simultaneous interpretation 
of HOMO and LUMO with IE and EA respectively. The comparison of KS gaps (difference between $\epsilon_{\mathrm{HOMO}}$ and 
$\epsilon_{\mathrm{LUMO}}$) with FG is beyond the physical consideration. We, however, note that for finite systems, 
one can always compute the FG from differences of ground-state energies (for anion, cation, neutral). We do not follow this 
approach; rather explore the credibility of our OT-RSH functionals for elucidation of frontier orbital energies in harmony 
with the quasi-hole and quasi-electron excitation. 

The comparison of (G)KS gap with FG depends solely on the choice of \emph{non-local}, orbital-specific operator within 
the (G)KS framework \citep{kronik12}. It has been shown that a judicious choice of non-local operator would diminish DD in the 
remaining potential \citep{kronik12} to such an extent that (G)KS gap would reflect FG accurately \citep{kronik12}. In RSH 
functionals, one can choose an optimal non-local operator by tuning $\gamma$ such that the residual DD in remaining potential 
would diminish. Therefore,
\begin{equation}
	E_{\mathrm{FG}} = \epsilon_{\mathrm{LUMO}}^{\gamma}-\epsilon_{\mathrm{HOMO}}^{\gamma},
\end{equation}
and immediately one can ask whether our tuning procedure is sufficient to choose an optimal non-local operator, such that it 
would satisfy Eqs.~[29,32] simultaneously. 

\begingroup                 
\squeezetable
\begin{table}     
\caption{\label{tab:table6} (G)KS gap vs experimental fundamental gap for selected atoms in eV.}
\resizebox{1.0\textwidth}{!}{
\begin{tabular} {l|ccccc|ccccccc|c}
\cline{1-14}
Atom & B3LYP & LC-BLYP & LC-BLYP$_{\mathrm{ot}}$ & CAM-B3LYP & CAM-B3LYP$_{\mathrm{ot}}$ & PBE$0$ & LC-PBE 
& LC-PBE$_{\mathrm{ot}}$ & CAM-PBE$0$ & CAM-PBE$0$$_{\mathrm{ot}}$ & LRC-$\omega$PBEh$^{\star}$ & 
LRC-$\omega$PBEh$^{\star}$$_{\mathrm{ot}}$ 
& Exp.\citep{johnson16} \\ 
\cline{1-14}
B  & 2.57 & 7.38 & 7.88 & 5.54 & 5.78 & 2.84 & 6.95 & 7.78 & 7.58 & 8.20 & 6.31 & 8.12 & 8.02 \\
C  & 3.19 & 8.34 & 9.72 & 6.55 & 7.19 & 3.59 & 7.69 & 9.21 & 8.86 & 10.00 & 7.16 & 9.84 & 10.00 \\
O  & 4.40 & 9.32 & 12.13 & 7.98 & 9.28 & 5.33 & 8.93 & 12.37 & 10.91 & 13.44 & 8.81 & 13.23 & 12.18 \\
Al & 1.48 & 5.57 & 5.63 & 3.89 & 3.91 & 1.77 & 5.45 & 5.66 & 5.66 & 5.81 & 4.95 & 6.02 & 5.56 \\
Si & 1.75  & 6.49 & 6.59 & 4.59 & 4.63 & 2.08 & 6.20 & 6.55 & 6.62 & 6.88 & 5.55 & 6.82 & 6.76 \\
P  & 4.07 & 9.11 & 9.61 & 7.14 & 7.37 & 4.96 & 9.29 & 10.11 & 9.91 & 10.52 & 8.64 & 10.44 & 9.74 \\

S  & 2.20 & 7.32 & 8.14 & 5.38 & 5.76 & 2.83 & 7.14 & 8.34 & 7.89 & 8.78 & 6.45 & 8.69 & 8.34 \\
Cl & 2.46 & 7.78 & 9.13 & 5.87 & 6.49 & 3.10 & 7.42 & 9.22 & 8.48 & 9.82 & 6.78 & 9.70 & 9.36 \\
Ga & 1.49 & 5.58 & 5.63 & 3.89 & 3.92 & 1.79 & 5.48 & 5.69 & 5.69 & 5.84 & 4.98 & 5.81 & 5.57 \\
Ge & 1.64 & 6.25 & 6.33 & 4.39 & 4.43 & 1.97 & 6.02 & 6.33 & 6.38 & 6.62 & 5.39 & 6.56 & 6.67 \\
As & 3.74 & 8.60 & 8.89 & 6.67 & 6.79 & 4.60 & 8.83 & 9.38 & 9.34 & 9.75 & 8.19 & 9.68 & 9.02 \\
Se & 1.92 & 6.90 & 7.39 & 4.96 & 5.18 & 2.50 & 6.78 & 7.59 & 7.35 & 7.96 & 6.08 & 7.89 & 7.73 \\

Br & 2.04 & 7.24 & 7.83 & 5.26 & 5.54 & 2.61 & 6.96 & 7.93 & 7.72 & 8.45 & 6.26 & 8.35 & 8.32 \\
\cline{1-14}
MAE & 5.72 & 0.88 & 0.20 & 2.70 & 2.38 & 5.18 & 1.09 & 0.26 & 0.49 & 0.38 & 1.67 & 0.34 &  \\
ME & 5.72 & 0.88 & 0.18 & 2.70 & 2.38 & 5.18 & 1.09 & 0.09 & 0.38 & --0.37 & 1.67 & --0.30 &  \\
$\Upsilon$ &  &  & 4.40 &  & 1.13 &  &  & 4.19 &  & 1.29 &  & 4.91 & \\
\cline{1-14}
\end{tabular}}
\begin{tabbing}
$^{\dagger}$$\Upsilon$: Ratio between MAE value of RSH and OT-RSH. 
\end{tabbing}
\end{table}
\endgroup

Before getting into FG, it is important to make an analysis between (G)KS LUMO energy and EA computed from the energy difference 
between two successive SCF calculation on neutral atom and anion. Table~V displays the calculated relative error in EA, 
$\Delta EA$ as, 
\begin{equation}
\Delta \mathrm{EA} = |(-\epsilon_{\mathrm{LUMO}^{\gamma}, \mathrm{Neutral}})-(E_{\mathrm{Neutral}}^{\gamma}- 
E_{\mathrm{Anion}}^{\gamma})|,
\end{equation}
for same set of atoms of Table~III. The performance of OT-RSH functionals is reflected through respective MAE values. Again, 
LRC-$\omega$PBEh$^{\star}$0$_{\mathrm{ot}}$ along with CAM-PBE0$_{\mathrm{ot}}$ show superiority in reducing $\Delta EA$ than other 
functionals. On the basis of MAE, one can arrange five functionals in descending order of performance as: 
CAM-PBE0$_{\mathrm{ot}}$, LRC-$\omega$PBEh$^{\star}$0$_{\mathrm{ot}}$, LC-PBE$_{\mathrm{ot}}$, LC-BLYP$_{\mathrm{ot}}$, 
CAM-B3LYP$_{\mathrm{ot}}$. Surprisingly, the $\Delta \mathrm{EAs}$ for OT-RSH functionals from B3LYP block is larger than PBE0 
block. It may be due to a better compatibility of latter block XC's with the remaining potential. In any case, the performance of 
OT-RSH hybrid functionals is improved from  respective RSH functionals for all the atoms in both blocks. 

Now, the computed (G)KS gap along with experimental values \citep{johnson16} are produced in Table~VI for all functionals for 
the same atoms of previous table. A closer look at both MAE and ME reveals that their values are close to each other for all 
functionals except LC-PBE$_{\mathrm{ot}}$. Moreover, the ME for CAM-PBE$0_{\mathrm{ot}}$ and 
LRC-$\omega$PBEh$^{\star}$$_{\mathrm{ot}}$ are negative, indicating that the systematic underestimation of FG of atoms does not 
happen most of the cases here, so nicely. One sees that,  LRC-$\omega$PBEh$^{\star}$$_{\mathrm{ot}}$ has the largest values of 
$4.91$ i.e., it reduces the error $4.91$ times more relative to that of LRC-$\omega$PBEh$^{\star}$. The five functionals in 
descending order are as follows: LRC-$\omega$PBEh$^{\star}$$_{\mathrm{ot}}$, LC-PBE$_{\mathrm{ot}}$, LC-BLYP$_{\mathrm{ot}}$, 
CAM-PBE$0_{\mathrm{ot}}$, CAM-B3LYP$_{\mathrm{ot}}$. On the other hand, if we compare them with MAE, then OT-RSH (LC) 
functionals seem to perform better than OT-RSH (LRC) and OT-RSH (CAM). As found in IV(A), here also, CAM-PBE0 appears 
to execute better than CAM-B3LYP, possibly for a similar reason, as delineated there. In any case, however, the 
accuracy of OT-RSH functionals is found to be greater than that of RSH for all species. Among them, 
LRC-$\omega$PBEh$^{\star}$$_{\mathrm{ot}}$ with MAE $= 0.34$ and LC-BLYP$_{\mathrm{ot}}$ with MAE $= 0.20$ exhibit excellent 
performance, which is very close to the recently published results using OT-BNL functional with a mean absolute deviation of 
$0.1$ eV \citep{stein10}. Here, the performance of OT-RSH functionals is limited by use of pseudo-valence basis and SR 
LDA/GGA exchange. Therefore, there is a strong possibility of further improvement using all-electron basis set, and 
incorporating modern SR exchange functionals. 

\begin{figure}             
\centering
\begin{minipage}[c]{0.30\textwidth}\centering
\includegraphics[scale=0.30]{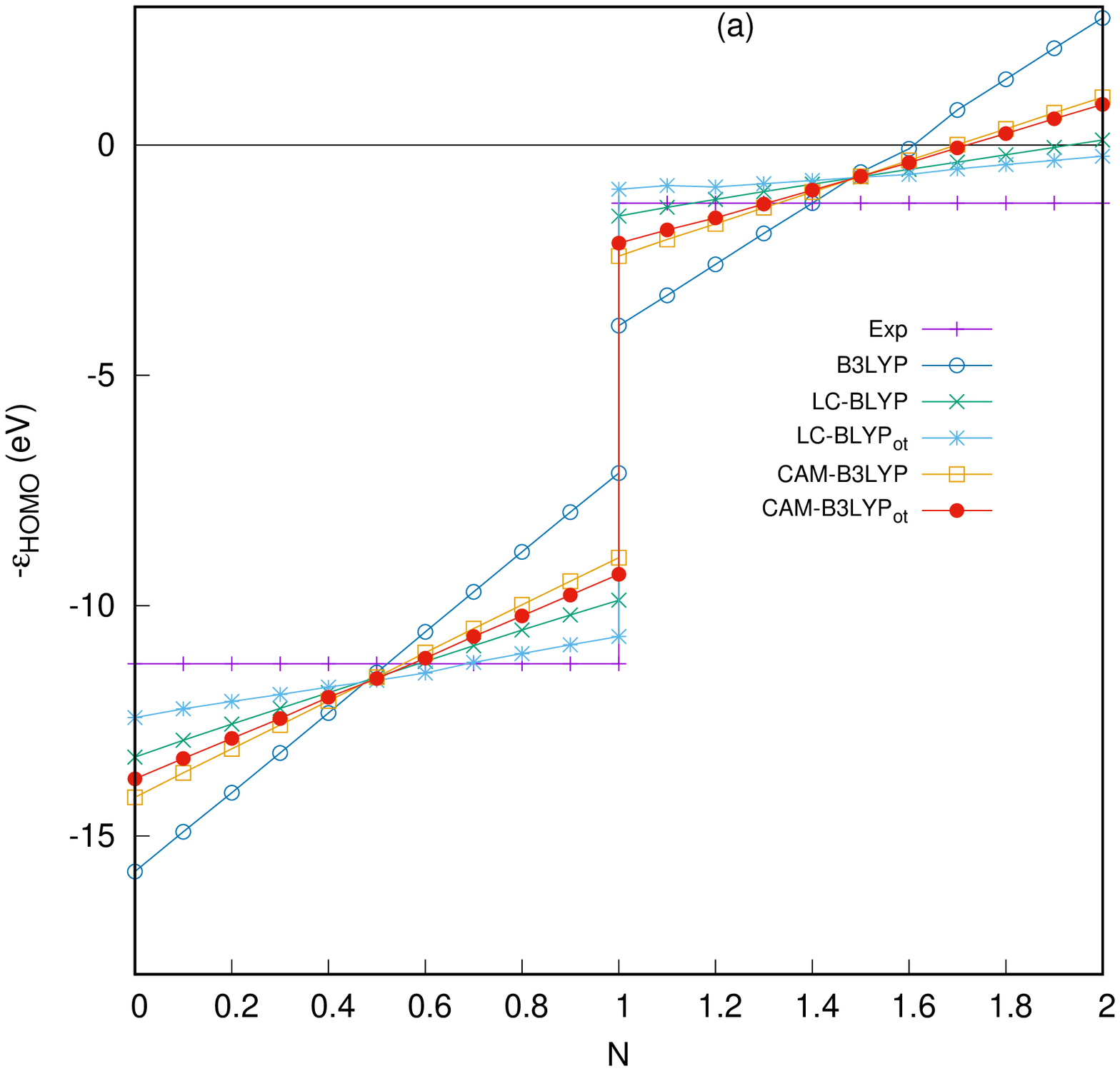}
\end{minipage}\hspace{0.1in}
\begin{minipage}[c]{0.30\textwidth}\centering
\includegraphics[scale=0.30]{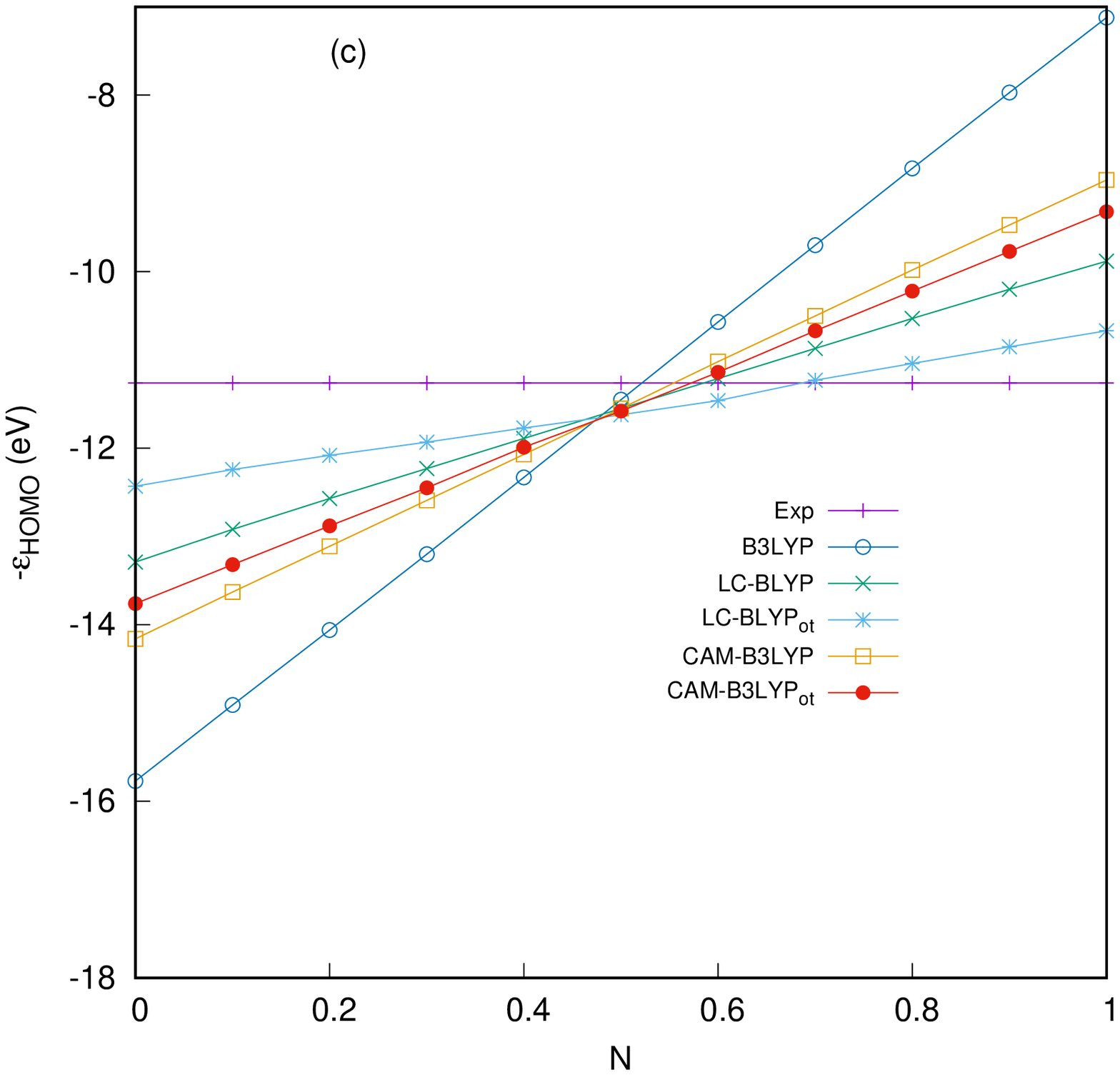}
\end{minipage}\hspace{0.1in}
\begin{minipage}[c]{0.30\textwidth}\centering
\includegraphics[scale=0.30]{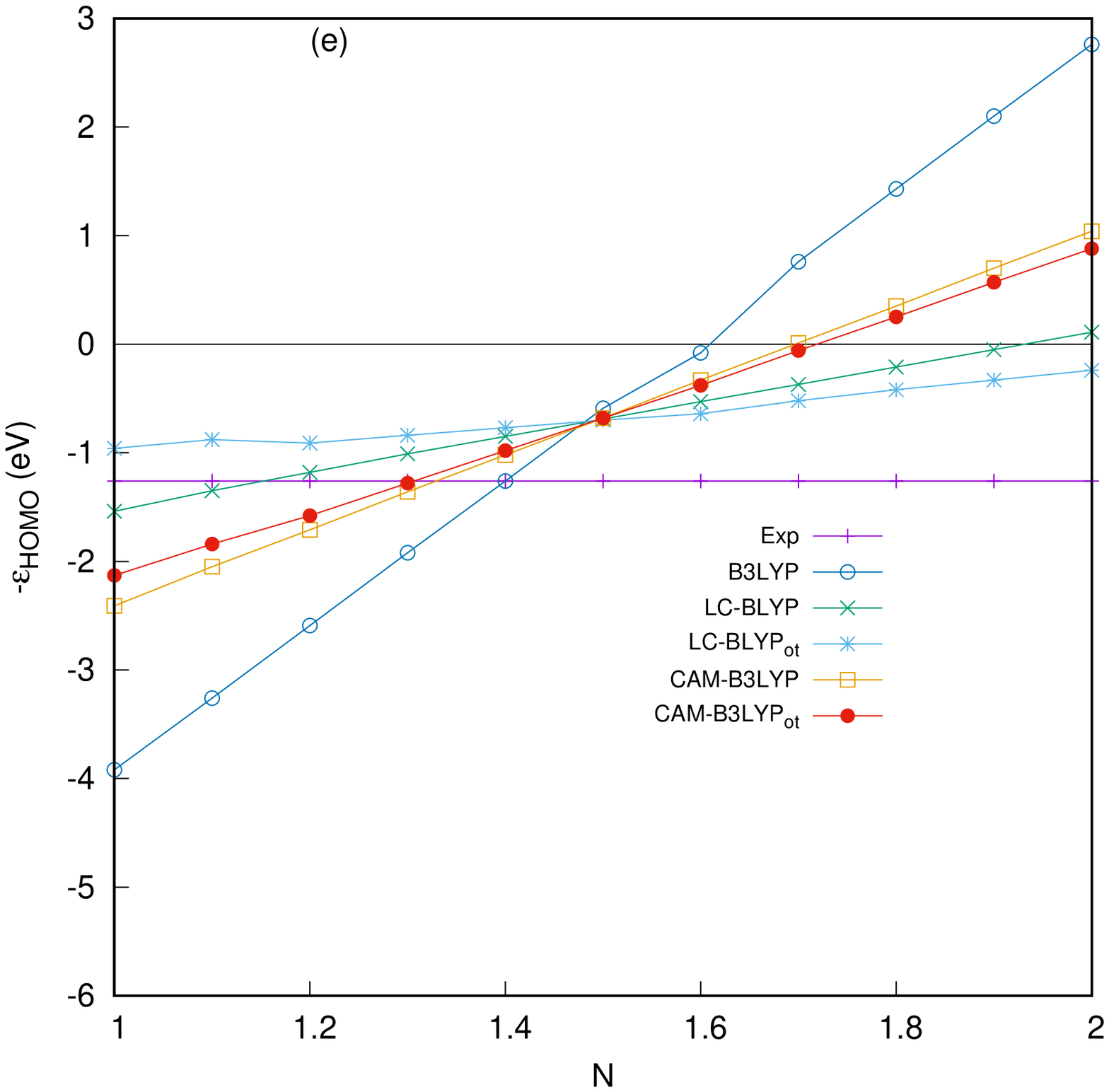}
\end{minipage}\hspace{0.1in}
\\[1pt]
\centering
\begin{minipage}[c]{0.30\textwidth}\centering
\includegraphics[scale=0.30]{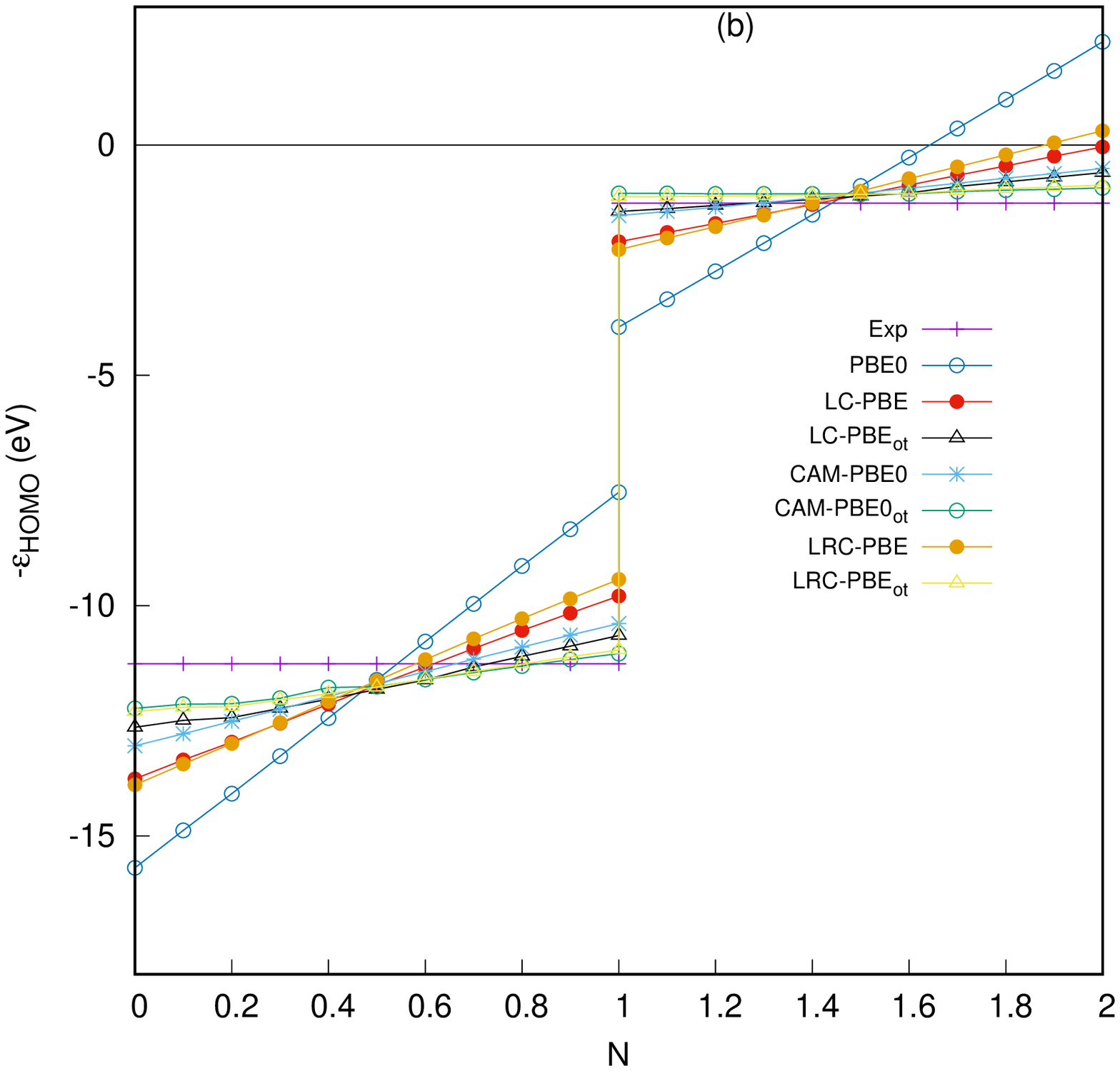}
\end{minipage}\hspace{0.1in}
\begin{minipage}[c]{0.30\textwidth}\centering
\includegraphics[scale=0.30]{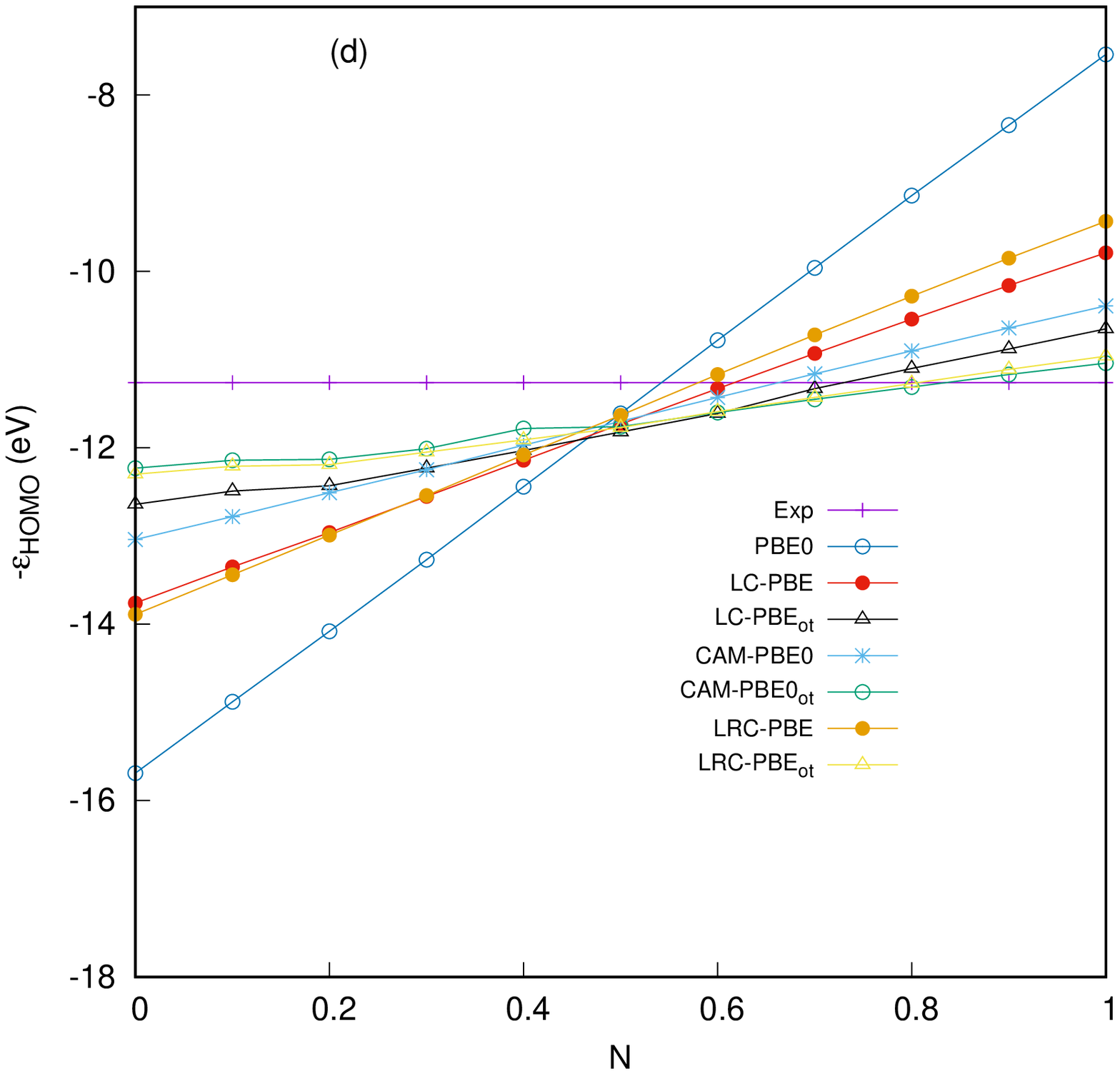}
\end{minipage}\hspace{0.1in}
\begin{minipage}[c]{0.30\textwidth}\centering
\includegraphics[scale=0.30]{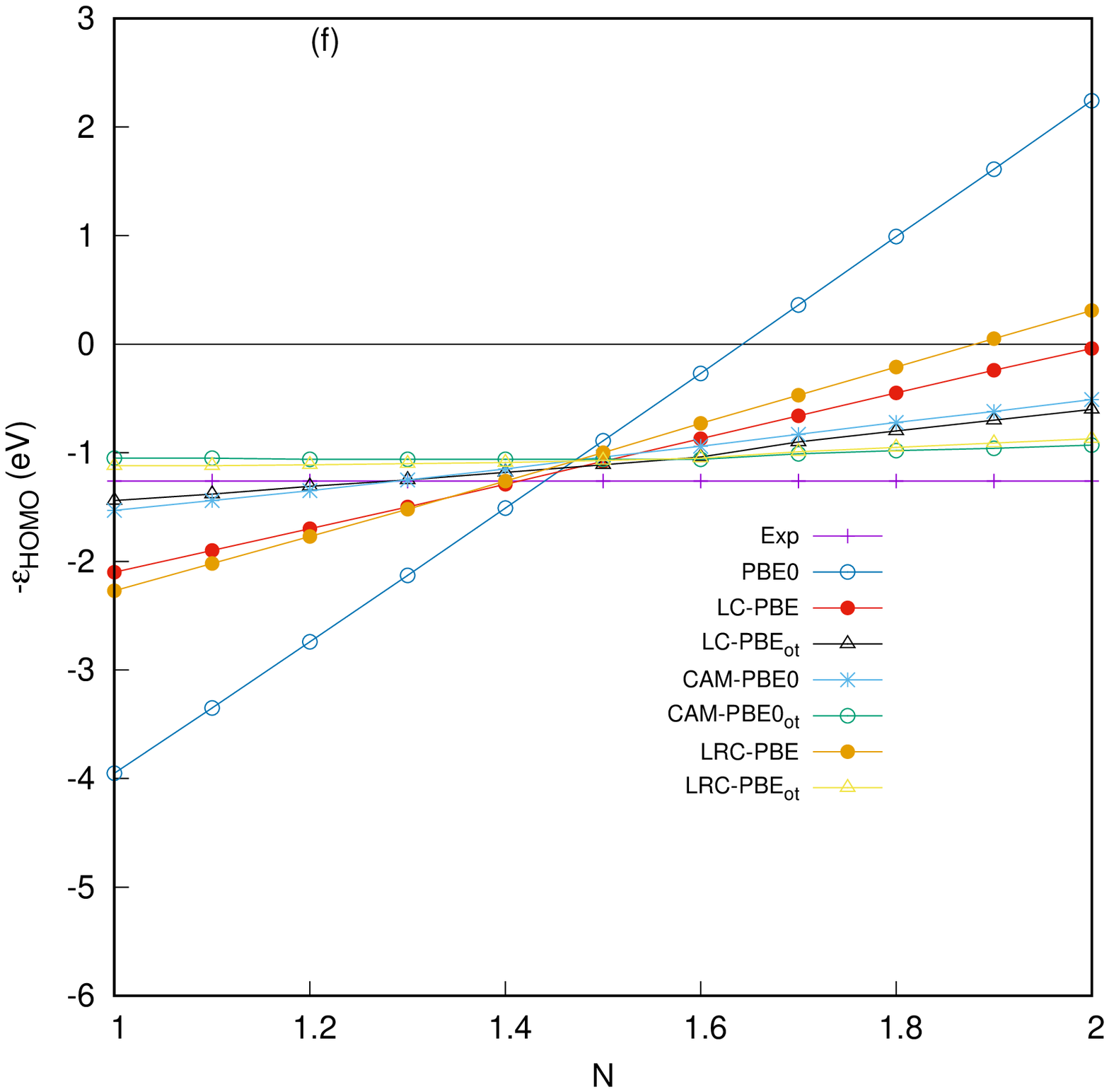}
\end{minipage}\hspace{0.1in}
\caption[optional]{Performance of different functionals on fractional occupation in C atom. The upper panel shows 
(a) highest occupied energy of C atom as a fraction of occupied p-electron number with $0 \leq  N \leq 2$, (c) $0 \leq N 
\leq 1$, and (e) $1 \leq N \leq 2$ for B3LYP block functionals. The bottom panels (b), (d) and (f) correspond to PBE0 block 
functionals.}
\end{figure}

\subsection{Fractional occupation in atoms}
An important challenge in DFT is a proper description of fractional charge systems. This was first introduced in
\citep{mori06} in relation to the exhibition of convexity in the curve of total energy $E(M+\delta)$ as function 
of fractional number of electron, $\delta$. According to the generalization of ground-state energy to systems with 
fractional number of electrons, the exact behavior of energy upon $\delta$ should be a straight line connecting 
the values at integer, the so-called PWL condition \citep{perdew82}. It can be defined as,
\begin{equation}
\begin{rcases}
E_{\mathrm{frac}}(N)=E(N)-E_{\mathrm{PWL}}(N), \quad N=M+\delta, \\
E_{\mathrm{PWL}}(N)=(1-\delta)E(M)+\delta E(M+1), \quad 0 \leq \delta \leq 1, \\
E_{\mathrm{PWL}}(N)=(1+\delta)E(M)-\delta E(M-1), \quad -1 \leq \delta \leq 0, \\
\end{rcases}
\end{equation}
where $E(N)$ and $E_{\mathrm{PWL}}(N)$ define the energy and PWL interpolation of energy for fractional number of 
electrons respectively. Actually, the value of $E_{\mathrm{frac}}$ provides a measure of deviation from PWL 
behavior. If $E_{\mathrm{frac}}  < 0$, the curve is convex and it is concave for $E_{\mathrm{frac}} > 0$. All the 
familiar DFAs face certain difficulties, giving smooth convex curve, whereas exact exchange shows an opposite 
trend. It has been found that there is a sign of improvement in the description of fractional number of electron 
systems with RSH functionals \citep{mori06} which combine these two ingredients in the respective inter-electronic 
regions. But still, they do not portray the PWL feature properly, sometimes quite convex. Furthermore, there is a 
great sensitivity of latter to the value of $\gamma$, besides other effects. Therefore, immediately one can ask 
whether our tuning procedure is sufficient to maintain PWL condition by satisfying Eq.~[34] along with Eqs.~[29,32] 
simultaneously. 

\begin{figure}             
\centering
\begin{minipage}[c]{0.30\textwidth}\centering
\includegraphics[scale=0.45]{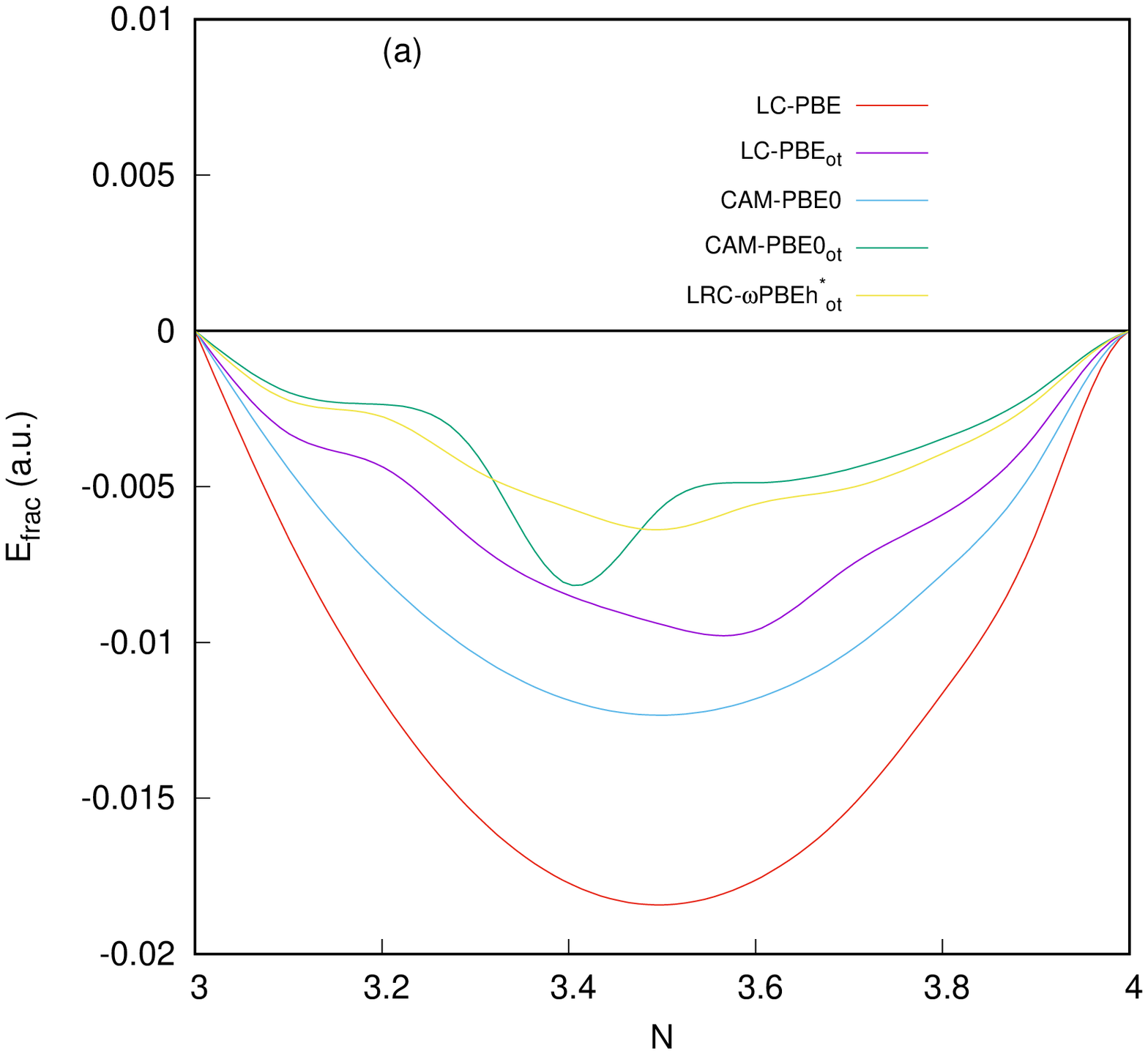}
\end{minipage}\hspace{1.5in}
\begin{minipage}[c]{0.30\textwidth}\centering
\includegraphics[scale=0.45]{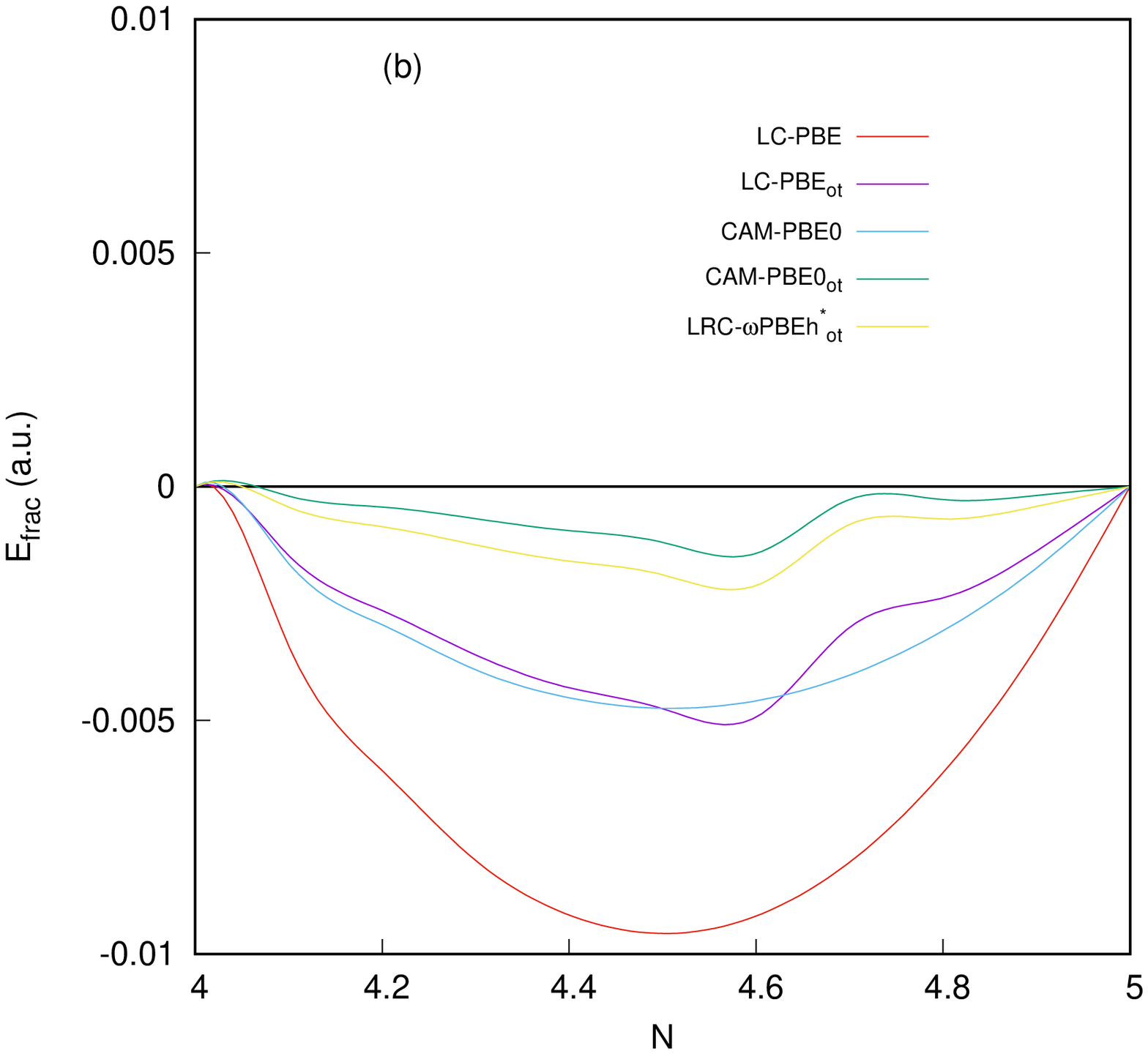}
\end{minipage}\hspace{1.0in}
\caption[optional]{Performance of different functionals on fractional occupation in C atom. The upper panel shows 
	(a) $E_{\text{frac}}$ of C atom as a fraction of total number of valence electrons with $3 \leq  N \leq 4$, and 
	(b) $4 \leq N \leq 5$ for PBE0 block functionals.}
\end{figure}

To put things in perspective, we consider Janak's theorem \citep{janak78}, and accordingly, each occupied eigenvalue 
is given by the derivative of energy with respect to its occupation number. Applying this rule to the (G)KS HOMO, 
the variation of $\epsilon_{\mathrm{HOMO}}$ with respect to fractional occupation number should be straight line. 
Then, there is finite jump due to presence of DD at the integer point, and again the variation of occupied 
$\epsilon_{\mathrm{LUMO}}$ with respect to the fractional occupation number should be straight line up to the next 
integer point. In order to do that we incorporate the fractional occupation number $n_i$ in the expression of 
density as,
\begin{equation}
\rho(\rvec)=\sum_i n_i|\phi_i(\rvec)|^2, n_i 
\begin{cases}
0, & i>i_{max} \\
\delta, & i=i_{max} \\
1, & i<i_{max}
\end{cases} 
\end{equation}
where $i_{max}$ corresponds to HOMO level. We assume no degeneracy in HOMO level, and remove the spin indices for simplicity. 
Here, we consider C atom as a specimen case, to demonstrate the relative performance of OT-RSH functionals. In an exact 
scenario, the behavior in the range $-1 \le \delta \le 0$ $(0 \le N \le 1)$ is obtained from experimental IE, while in 
$0 \le \delta \le 1$ $(1 \le N \le 2)$, from experimental EA. In Fig.~1, we demonstrate the performance of RSH functionals 
in C atom. The upper panel contains B3LYP block functionals in three respective regions: $0 \leq  N \leq 2$ (Fig.~1a), 
$0 \leq N \leq 1$ (Fig~1c), and $1 \leq N \leq 2$ (Fig.~1e). From Fig.~1a, one can observe the pattern across the whole 
region along with a step-like feature at integer point for all functionals except B3LYP and PBE0. The OT-RSH functionals 
clearly perform better than the respective RSH functionals. A closer look at Fig.~1c and 1e uncovers that the 
LC-BLYP$_{\mathrm{ot}}$ is in close proximity with the straight-line behavior in both regions, and it shows a much better 
performance in $1 \leq N \leq 2$. Furthermore, the performance of CAM-B3LYP$_{\mathrm{ot}}$ is not so pronounced than 
CAM-B3LYP; and not even of LC-BLYP. 

\begin{figure}             
\centering
\begin{minipage}[c]{0.30\textwidth}\centering
\includegraphics[scale=0.45]{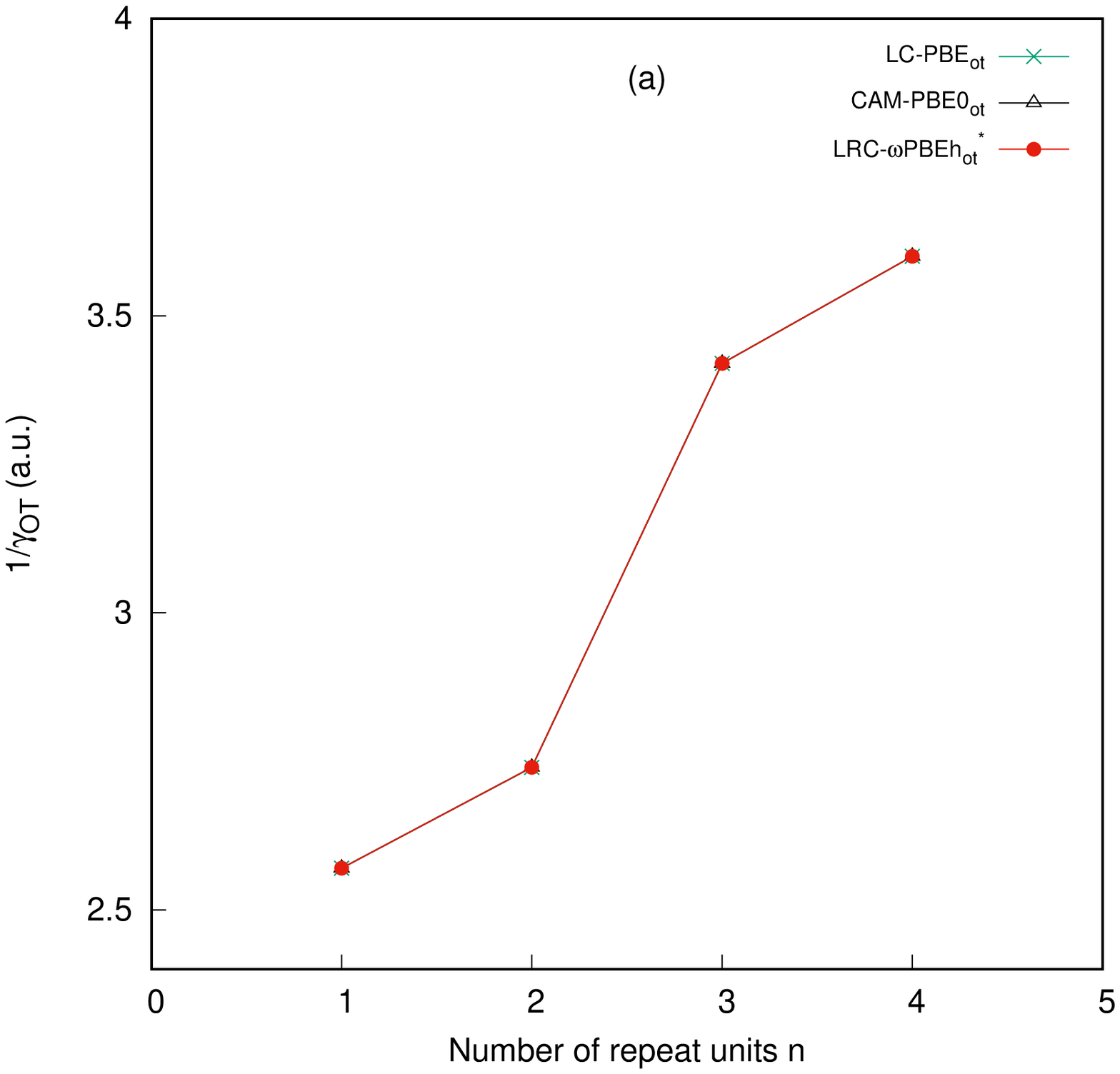}
\end{minipage}\hspace{1.5in}
\begin{minipage}[c]{0.30\textwidth}\centering
\includegraphics[scale=0.45]{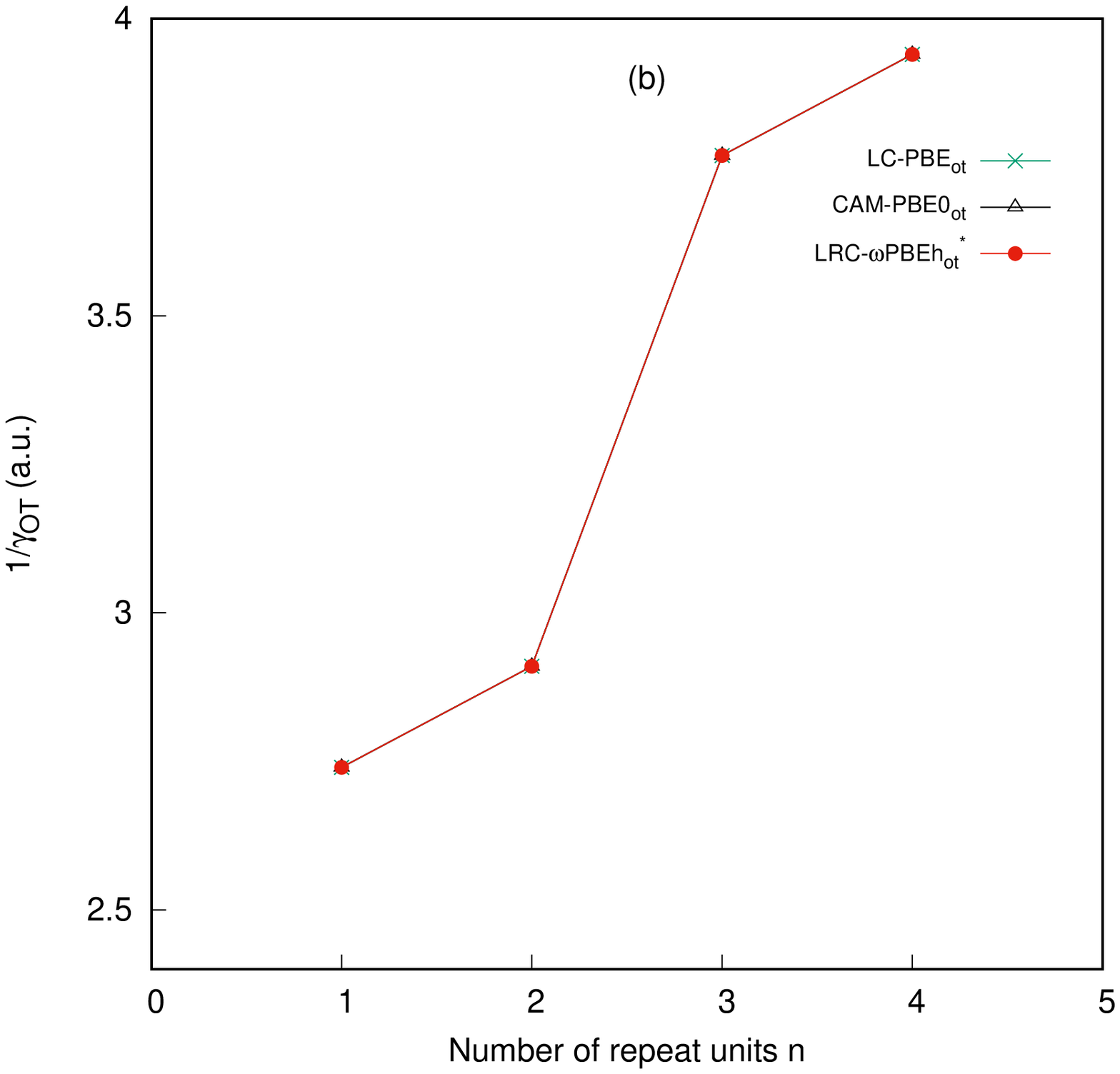}
\end{minipage}\hspace{1.0in}
	\caption[optional]{The characteristic length of SR/LR separation ($1/\gamma_{\mathrm{OT}}$) for 
	(a) linear polyene chains (C$_{2n}$H$_{2n+2}$) (b) linear alkane chains (C$_{2n}$H$_{4n+2}$), as a
	function of $n$.}
\end{figure}

On the other hand, the lower block of Fig.~1 corresponds to the PBE0 functional, in three respective regions: 
$0 \leq  N \leq 2$ (Fig.~1b), $0 \leq N \leq 1$ (Fig~1d), and $1 \leq N \leq 2$ (Fig.~1f). Here also, OT-RSH functionals 
fare better than conventional RSH. From a glance at Fig.~1d and 1f, it follows that 
LRC-$\omega$PBEh$^{\star}$$_{\mathrm{ot}}$ and CAM-PBE0$_{\mathrm{ot}}$ behave in a quite similar manner with each other. 
They also behave very close to experiment in the region $1 \leq N \leq 2$ with a small overall positive shift in energy. 
From a comparison of two blocks, it appears OT-RSHs (PBE0) appears to be significantly better than OT-RSHs (B3LYP); more 
specifically the CAM-PBE0$_{\mathrm{ot}}$ fares much better than that of CAM-B3LYP$_{\mathrm{ot}}$. Based on all these 
facts, now one can arrange the five OT functionals in descending order of performance as: 
LRC-$\omega$PBEh$^{\star}$$_{\mathrm{ot}} \approx$ CAM-PBE0$_{\mathrm{ot}} >$ 
LC-PBE$_{\mathrm{ot}}\approx$ LC-BLYP$_{\mathrm{ot}} >$ CAM-B3LYP$_{\mathrm{ot}}$. 

\color{red}
Further, we have also presented the variation of $E_{\text{frac}}$ according to Eq.~[34] along with Eqs.~[29,32]. Once again,  
C is chosen as a specimen case, to illustrate the relative performance of OT-RSH functionals on PWL behaviour. In an exact 
scenario, the nature in the range ($3 \ge N \ge 4$) and ($4 \ge N \ge 5$) should be straight line. Figure~2 
offers this for RSH functionals from PBE0 block; all the five functionals record convex nature, with LC-PBE showing maximum. 
In keeping with Fig.~1, here also OT-RSH functionals fare better than the conventional RSH. It also appears that, 
LRC-$\omega$PBEh$^{\star}_{\text{ot}}$ and CAM-PBE0$_{\text{ot}}$ show some similarity between them. These two turn out to  
be the closest ones near the straight line in region ($4 \ge N \ge 5$) with much reduced convex deviation. 
This establishes the fact that the current tuning procedure is sufficient to maintain PWL condition with a considerably small 
convex deviation. 

\color{black}

\begin{figure}             
\centering
\includegraphics[scale=0.8]{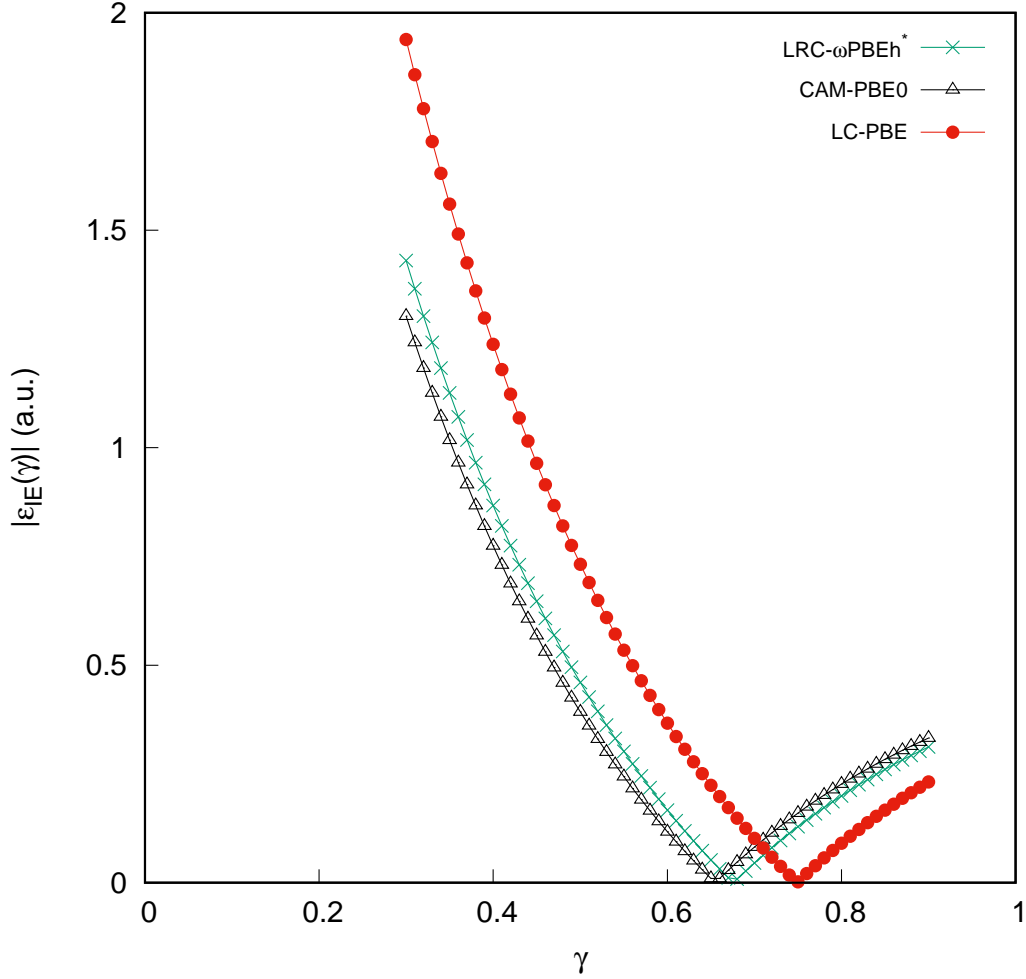}
\hspace{1.0in}
	\caption[optional]{The difference in ionization energies ($|\varepsilon_{\text{IE}}(\gamma)|$) for 
	carbon atom as a function of $\gamma$.}
\end{figure}

\subsection{Short polyenes and alkane chains}
It is known that molecules of similar size can still possess a $\gamma_{\mathrm{OT}}$ that is quite different, and
the extent of conjugation plays an important role in determining this, besides size \citep{korzdorfer11}. 
To make a direct comparison in such systems, here we pick polyenes, C$_{2n}$H$_{2n+2}$, and alkane chains, 
C$_{2n}$H$_{4n+2}$, with $n=1-4$, from \citep{korzdorfer11}. In this occasion, the repeat unit ($n$) is defined in a 
way such that they are consistent with each other. To pursue this, we plot 
the characteristic length of SR/LR separation i.e., $1/\gamma_{\mathrm{OT}}$ as defined in \citep{korzdorfer11}, 
as a function of number of repeat unit, in panels (a), (b) of Fig.~3, for polyenes and alkane chains respectively. Only 
three functionals from PBE0 block are adopted, as the choice of semi-local functionals in determining $\gamma_{\text{OT}}$ 
is completely negligible. From these, a very strong dependence of $1/\gamma_{\mathrm{OT}}$ on chain length is observed, 
for all three functionals, having identical slope. It is expected, as our self-consistent systematic optimization 
procedure is based on size dependency principle. To explore the role of conjugation in determining 
$\gamma_{\mathrm{OT}}$, a comparison between polyenes versus alkane chains is very important, and thus undertaken. It 
reveals that, the evolution of $1/\gamma_{\text{OT}}$ is linear and similar in both cases. But the 
value in alkane chains is larger than those for polyenes. Thus, it reaffirms the crucial role played by the extent of 
conjugation in such a scenario. Therefore, it is consistent with the analysis of \citep{korzdorfer11}. A more elaborate 
comparison for larger $n$ may be considered in future; however this basic analysis augurs well for the optimization procedure 
proposed here.

\begingroup                 
\squeezetable
\begin{table}     
	\caption{\label{tab:table7}Present (PR1) vs conventional (PR2) optimally tuned IEs for 
	atoms, in eV.}
\resizebox{1.0\textwidth}{!}{
\begin{tabular} {l|ccccccc|ccccccc|c}
\cline{1-16}
	& \multicolumn{6}{c}{PR1} & & \multicolumn{6}{c}{PR2} & & \\ 
\cline{1-16}
	Atoms & LC-PBE$_{\text{OT}}$ & $\gamma_{\text{OT}}$ & CAM-PBE$0_{\text{OT}}$ & $\gamma_{\text{OT}}$ & 
	LRC-$\omega$PBEh$_{\text{OT}}^{\star}$ &  $\gamma_{\text{OT}}$ & & LC-PBE$_{\text{OT}}$ & $\gamma_{\text{OT}}$ & 
	CAM-PBE$0_{\text{OT}}$ & $\gamma_{\text{OT}}$ & LRC-$\omega$PBEh$_{\text{OT}}^{\star}$ & $\gamma_{\text{OT}}$ & 
	& Exp.\citep{johnson16} \\ 
\cline{1-16}
	Be  & 8.67 & 0.32 & 8.78 & 0.32 & 8.75 & 0.32 &  & 9.08 & 0.75 & 9.08 & 0.70 & 9.08 & 0.72 & & 9.32 \\
	B  & 7.93 & 0.39 & 8.16 & 0.39 & 8.11 & 0.39 &  & 8.74 & 0.80 & 8.73 & 0.73 & 8.73 & 0.75 & & 8.02 \\
	C  & 10.65 & 0.42 & 11.04 & 0.42 & 10.96 & 0.42 &  & 11.82 & 0.75 & 11.76 & 0.65 & 11.79 & 0.68 & & 11.26 \\
	N  & 13.79 & 0.49 & 14.30 & 0.49 & 14.20 & 0.49 &  & 15.18 & 0.82 & 15.08 & 0.70 & 15.11 & 0.73 & & 14.53 \\
	O  & 13.00 & 0.53 & 13.50 & 0.53 & 13.40 & 0.53 & & 13.73 & 0.69 & 13.66 & 0.57 & 13.69 & 0.6 & & 13.62 \\
	Si & 8.01  & 0.34 & 8.22 & 0.34 & 8.18 & 0.34 & & 8.69 & 0.69 & 8.68 & 0.62 & 8.68 & 0.64 & & 8.15 \\
	P  & 10.13 & 0.39 & 10.40 & 0.39 & 10.35 & 0.39 & & 10.91 & 0.68 & 10.90 & 0.61 & 10.91 & 0.63 & & 10.49 \\
	S  & 10.00 & 0.42 & 10.24 & 0.42 & 10.19 & 0.42 & & 9.13 & 0.48 & 9.13 & 0.42 & 9.12 & 0.43 & & 10.36 \\
	Cl & 12.57 & 0.45 & 12.92 & 0.45 & 12.85 & 0.45 & & 13.27 & 0.64 & 13.27 & 0.56 & 13.27 & 0.58 & & 12.97 \\
	Ge & 7.62 & 0.34 & 7.79 & 0.34 & 7.76 & 0.34 & & 8.20 & 0.68 & 8.19 & 0.62 & 8.19 & 0.63 & & 7.90 \\
	As & 9.57 & 0.36 & 9.82 & 0.36 & 9.77 & 0.36 & & 10.31 & 0.66 & 10.30 & 0.59 & 10.31 & 0.61 & & 9.82 \\
	Se & 9.28 & 0.39 & 9.49 & 0.39 & 9.47 & 0.39 & & 9.71 & 0.54 & 9.69 & 0.47 & 9.70 & 0.49 & & 9.75 \\
	Br & 11.30 & 0.39 & 11.60 & 0.39 & 11.54 & 0.39 & & 11.85 & 0.54 & 11.95 & 0.51 & 11.94 & 0.52 & & 11.81 \\
\cline{1-16}
	MAE & 0.41 &  & 0.15 &  & 0.20 &  & & 0.43 &  & 0.41 &  & 0.42 &  & &  \\
	ME & 0.41 &  & 0.15 &  & 0.19 &  & & --0.31 &  & --0.29 &  & --0.30 & & & \\
\cline{1-16}
\end{tabular}}
\end{table}
\endgroup

\subsection{A comparison with traditional optimal tuning}
At last, a comparison in the accuracy between present results and conventional OT procedure 
\citep{livshits07, stein10, tamblyn14, kronik12}, based on Koopmans' theorem, would be worthwhile. 
Thus we have examined the IEs of same set of atoms from Table~III. In the latter scheme, $\gamma$ is tuned in such a way 
that the HOMO eigenvalue of neutral system equals IE, with the latter determined as a difference 
in ground-state energy of neutral ($E(\gamma,N)$) and cationic ($E(\gamma,N-1)$) species. Consequently, 
the IE-optimized $\gamma$ can be obtained by minimizing,  
\begin{equation}
	\varepsilon_{\text{IE}}(\gamma)=|\epsilon_{\text{HOMO}}^{\gamma}-E(\gamma,N)-E(\gamma,N-1)|.
\end{equation}
As a specimen case, the implementation of this method is presented in Fig.~4 considering C atom, using three PBE0 
block functionals. It is revealed that each plot of $\varepsilon_{\text{IE}}(\gamma)$ vs $\gamma$ finds a
unique minima for three functionals. This procedure is followed for all atoms, enabling one to estimate the respective 
$\gamma_{\text{OT}}$'s. The $\gamma_{\text{OT}}$ 
and its corresponding IE values are presented in Table~VII (PR2 columns). As expected, this procedure gives a unique 
value of $\gamma$ for each of them that depends strongly on the electronic structure, and to 
a lesser extent on the choice of semi-local functionals. The present estimates of $\gamma_{\text{OT}}$ along with 
respective IE's are provided in PR1 columns. Similar general observations as in PR2, also hold for PR1. 
It is noticed that $\gamma_{\text{OT}}$ is significantly lower in PR1, from PR2. 

\begingroup                 
\squeezetable
\begin{table}     
	\caption{\label{tab:table8}Present (PR1) vs conventional (PR2) optimally tuned IEs for 
	molecules, in eV.}
\resizebox{1.0\textwidth}{!}{
\begin{tabular} {l|ccccc|ccccc|c}
\cline{1-12}
	& \multicolumn{4}{c}{PR1} & & \multicolumn{4}{c}{PR2} & & \\ 
\cline{1-12}
	Molecules &LC-BLYP$_{\text{OT}}$ & $\gamma_{\text{OT}}$ & LC-PBE$_{\text{OT}}$  & $\gamma_{\text{OT}}$ 
	& &LC-BLYP$_{\text{OT}}$  & $\gamma_{\text{OT}}$ & LC-PBE$_{\text{OT}}$ & $\gamma_{\text{OT}}$ & 
	& Exp.\citep{johnson16} \\ 
\cline{1-12}
	N$_2$ & 14.81 &0.39  & 14.82 & 0.39 & & 16.37 & 0.64 & 16.43 & 0.65 & & 15.60 \\
Si$_2$  & 7.90 & 0.32 & 8.16 & 0.32 & & 7.35 & 0.43 & 8.29 & 0.44 & & 7.90 \\
P$_2$   & 10.50 & 0.36 & 10.69 & 0.36 & & 10.91 & 0.51 & 11.13 & 0.48 & & 10.62 \\
S$_2$    & 8.69 & 0.36 & 8.86 & 0.36 & & 10.13 & 0.52 & 9.48 & 0.51 & & 9.40 \\
Cl$_2$    & 11.78 & 0.42 & 11.99 & 0.42 & & 12.61 & 0.62 & 12.82 & 0.62 & & 11.48 \\
CO$_2$     & 13.58 & 0.39 & 13.59 & 0.39 & & 13.96 & 0.51 & 14.10 & 0.52 & & 14.10 \\
HCl     & 12.35 & 0.49 & 12.54 & 0.49 & & 12.90 & 0.58 & 13.06 & 0.57 & & 12.79 \\
CO     & 13.58 & 0.39 & 13.59 & 0.39 & & 14.28 & 0.50 & 14.24 & 0.49 & & 14.10 \\
H$_2$O     & 11.51 & 0.39 & 11.55 & 0.39 & & 12.59 & 0.55 & 12.64 & 0.55 & & 12.62 \\
H$_2$S     & 10.11 & 0.39 & 10.29 & 0.39 & & 10.60 & 0.51 & 10.71 & 0.49 & & 10.48 \\
C$_2$H$_2$     & 11.04 & 0.39 & 11.24 & 0.39 & & 11.32 & 0.46 & 11.49 & 0.45 & & 11.40 \\
NH$_3$     & 9.86 & 0.39 & 9.96 & 0.39 & & 10.48 & 0.49 & 10.58 & 0.49 & & 10.82 \\
PH$_3$     & 9.96 & 0.36 & 10.10 & 0.36 & & 10.36 & 0.46 & 10.43 & 0.44 & & 9.89 \\
CH$_4$     & 13.78 & 0.42 & 13.91 & 0.42 & & 14.17 & 0.49 & 14.29 & 0.49 & & 13.60 \\
CH$_3$Cl     & 11.04 & 0.39 & 11.20 & 0.39 & & 11.39 & 0.46 & 11.55 & 0.46 & & 11.29 \\
C$_2$H$_4$     & 10.18 & 0.36 & 10.41 & 0.36 & & 10.39 & 0.42 & 10.56 & 0.40 & & 10.51 \\
C$_2$H$_6$     & 11.95 & 0.36 & 12.06 & 0.36 & & 12.49 & 0.46 & 12.60 & 0.46 & & 11.99 \\
SiH$_4$     & 12.37 & 0.34 & 12.49 & 0.34 & & 12.96 & 0.46 & 13.12 & 0.47 & & 12.30 \\
Si$_2$H$_6$     & 10.59 & 0.32 & 10.75 & 0.32 & & 10.81 & 0.37 & 10.93 & 0.36 & & 10.53 \\
\cline{1-12}
	MAE & 0.38 &  & 0.36 &  & & 0.32 &  & 0.40 &  & &  \\
	ME & 0.31 &  & 0.17 &  & & --0.19 &  & --0.37 &  & &  \\
\cline{1-12}
\end{tabular}}
\end{table}
\endgroup

Let us now discuss the accuracy of these two methods, in terms of MAE and ME values collected at the bottom of the 
table. A glance reveals that, for all three functionals, these are identical in PR1, with all ME having ($+$)ve sign.  
In contrast, the ME values are found to be ($-$)ve for all functionals in case of conventional PR2 scheme. This indicates 
an underestimation and overestimation of IE from experiment, in PR1 and PR2 respectively. However, these error are systematic 
in nature. Now, the performance of these methods is reflected through respective MAE values. Though the MAE of 
LC-PBE$_{\text{OT}}$ in PR1 is very close to that of PR2, but the same in case of remaining two functionals  
is substantially lower (at least by a factor of 2) in PR1 from their PR2 counterparts, implying PR1 to be supposedly more 
accurate than PR2. 

\color{red}
Next, we also consider the IEs of same set of molecules from Table~IV (excluding NaCl) using present and conventional OT 
procedure as presented in Table~VII, for atoms. It is already mentioned that the auxiliary parameters ($\alpha, \beta, 
a^{\text{x,sr}}$) used in different functionals may have some sensitivity during self-consistent tuning process, and it 
is indeed important for molecular systems. To be consistent with this fact, here, two such functionals such as LC-BLYP 
and LC-PBE, which are devoid of auxiliary parameters, are used for an unbiased comparison. In the conventional OT scheme, 
we follow the same procedure as depicted in Fig.~4 for all molecules. The $\gamma_{\text{OT}}$ and its corresponding IE 
values are presented in Table~VIII (PR2 columns). As in atoms, this procedure gives a unique value of $\gamma$ for each 
of them that depends strongly on the electronic structure, and to a lesser extent on the choice of semi-local 
functionals. The present estimates of $\gamma_{\text{OT}}$ along with respective IE's are provided in PR1 columns. 
Similar general observations as in PR2, also hold for PR1. Further, we also found the same trend in $\gamma_{\text{OT}}$; 
which is usually lower in PR1, compared with PR2. 

A glance at MAE and ME at the bottom of the table reveals that, for two 
such functionals, these are quite close for PR1 and PR2 having ($+$)ve and ($-$)ve signs in MEs. This indicates 
an underestimation and overestimation of IE from experiment, in PR1 and PR2 respectively. However, these error are not 
systematic as found in case of atoms. Moreover, the MAE values signify that these two methods are reasonably close to each 
other in terms of accuracy. It is to be noted that $\gamma_{\text{OT}}$ values differ significantly (for the same system and 
functional) from the IP-tuning approach, which was also observed in case of atoms. But, it is consistent with the fact that 
if we follow carefully the ME values which are (+) and (-) signs for the present method and IP-tuning approach respectively. 

\color{black}
A few remarks may now be made before passing. The computational overhead of each step during optimization is at same 
level as in commonly used standard RSH functionals; the additional time is required only for grid optimization as discussed 
in Tables~I, II. The computation cost for each optimization step with respect to system size has already been mentioned in 
Sec.~II(B), and hence not repeated here. Moreover, if we consider $t_{\text{SCF}}$ to be the time required to complete one 
SCF cycle, $N_{\text{SCF}}$ the number of SCF iteration and $N_{\text{OP}}^{\text{PR1}}$ as total number of optimization 
steps, then the real time computational cost can be simply estimated as: $t_{\text{SCF}} \times N_{\text{SCF}} \times 
N_{\text{OP}}^{\text{PR1}}$. This equivalently holds for PR2 method. However, a real time difference can only come through 
the respective value of $N_{\text{OP}}^{\text{PR1}}$ and $N_{\text{OP}}^{\text{PR2}}$. Now the basic advantage of our method 
is its ease of implementation 
through a simple energy optimization, keeping the development from \emph{first principles} and this may offer considerable benefit 
for periodic systems, for which calculation on charged systems can be rather difficult. Moreover, it is 
sufficient to maintain Koopmans' theorem \citep{salzner09} and PWL conditions without imposing it during optimization. In future, it 
may be a viable alternative to popular \emph{ab initio} OT-RSH schemes \citep{livshits07, stein10, tamblyn14, kronik12}. It would 
be worthwhile to compare our results with the latter scheme in terms of accuracy and cost in a more elaborate fashion for other 
properties besides IE. Currently this is being pursued and may be communicated in future. 

\section{Future and Outlook}
We have demonstrated the feasibility and practicability of a self-consistent systematic optimization procedure for OT-RSH functionals 
from \emph{first principles}. This was applied for a host of atoms and molecules; properties derived from frontier orbital energies 
were offered, within a pseudopotential (G)KS-DFT framework. Besides these, their performances on 
fractional occupation of electron on HOMO were also presented. The accuracy of OT-RSH functionals is always improved than that of 
RSH functionals for all the species in case of both B3LYP and PBE0 series, in case of the properties derived from orbital energies. 
The success of this approach relies on accurate estimation of $\gamma_{\mathrm{OT}}$ based on size dependency principle. 

So far, we have taken in to account only the properties derived from frontier orbital energies. But, other properties derived from total 
energy of a system, may also improve the quality of these designed functionals. In this scenario, an extensive analysis is 
required to maintain the \emph{size consistency} on total energy. Again, the compatibility of $\gamma_{\mathrm{OT}}$ with auxiliary 
parameters present in RSH functionals should be taken into consideration during the optimization procedure. Furthermore, the relative 
performance of B3LYP block functionals seems to be less affected by our proposed scheme. This leads us to conclude that, there is a 
possibility of further improving the results, notably, by incorporating the recently developed SR GGA exchange functionals satisfying a 
set of prescribed constraints, and using better basis set from \emph{all-electron} calculations. In essence, we believe that the current 
approach may provide a useful practical framework for future development and application (such as static electric response properties, 
charge-transfer excitation, Rydberg excitation).

\section{Availability of data}
Data available on request from the authors.

\section{Acknowledgement}
AG is grateful to UGC for a senior research fellowship. AKR thankfully acknowledges funding from DST SERB, New Delhi, India 
(sanction order: CRG/2019/000293).
\bibliography{dft_ref.bib}

\begin{thebibliography}{85}
\expandafter\ifx\csname natexlab\endcsname\relax\def\natexlab#1{#1}\fi
\expandafter\ifx\csname bibnamefont\endcsname\relax
  \def\bibnamefont#1{#1}\fi
\expandafter\ifx\csname bibfnamefont\endcsname\relax
  \def\bibfnamefont#1{#1}\fi
\expandafter\ifx\csname citenamefont\endcsname\relax
  \def\citenamefont#1{#1}\fi
\expandafter\ifx\csname url\endcsname\relax
  \def\url#1{\texttt{#1}}\fi
\expandafter\ifx\csname urlprefix\endcsname\relax\def\urlprefix{URL }\fi
\providecommand{\bibinfo}[2]{#2}
\providecommand{\eprint}[2][]{\url{#2}}

\bibitem[{\citenamefont{Hohenberg and Kohn}(1964)}]{hohenberg64}
\bibinfo{author}{\bibfnamefont{P.}~\bibnamefont{Hohenberg}} \bibnamefont{and}
  \bibinfo{author}{\bibfnamefont{W.}~\bibnamefont{Kohn}},
  \bibinfo{journal}{Phys.~Rev.} \textbf{\bibinfo{volume}{136}},
  \bibinfo{pages}{B864} (\bibinfo{year}{1964}).

\bibitem[{\citenamefont{Cohen et~al.}(2011)\citenamefont{Cohen,
  Mori-S{\'a}nchez, and Yang}}]{cohen11}
\bibinfo{author}{\bibfnamefont{A.~J.} \bibnamefont{Cohen}},
  \bibinfo{author}{\bibfnamefont{P.}~\bibnamefont{Mori-S{\'a}nchez}},
  \bibnamefont{and} \bibinfo{author}{\bibfnamefont{W.}~\bibnamefont{Yang}},
  \bibinfo{journal}{Chem.~Rev.} \textbf{\bibinfo{volume}{112}},
  \bibinfo{pages}{289} (\bibinfo{year}{2011}).

\bibitem[{\citenamefont{Burke}(2012)}]{burke12}
\bibinfo{author}{\bibfnamefont{K.}~\bibnamefont{Burke}},
  \bibinfo{journal}{J.~Chem.~Phys.} \textbf{\bibinfo{volume}{136}},
  \bibinfo{pages}{150901} (\bibinfo{year}{2012}).

\bibitem[{\citenamefont{Becke}(2014)}]{becke14}
\bibinfo{author}{\bibfnamefont{A.~D.} \bibnamefont{Becke}},
  \bibinfo{journal}{J.~Chem.~Phys.} \textbf{\bibinfo{volume}{140}},
  \bibinfo{pages}{18A301} (\bibinfo{year}{2014}).

\bibitem[{\citenamefont{Jones}(2015)}]{jones15}
\bibinfo{author}{\bibfnamefont{R.~O.} \bibnamefont{Jones}},
  \bibinfo{journal}{Rev.~Mod.~Phys.} \textbf{\bibinfo{volume}{87}},
  \bibinfo{pages}{897} (\bibinfo{year}{2015}).

\bibitem[{\citenamefont{Kohn and Sham}(1965)}]{kohn65}
\bibinfo{author}{\bibfnamefont{W.}~\bibnamefont{Kohn}} \bibnamefont{and}
  \bibinfo{author}{\bibfnamefont{L.~J.} \bibnamefont{Sham}},
  \bibinfo{journal}{Phys.~Rev.} \textbf{\bibinfo{volume}{140}},
  \bibinfo{pages}{A1133} (\bibinfo{year}{1965}).

\bibitem[{\citenamefont{Seidl et~al.}(1996)\citenamefont{Seidl, G{\"o}rling,
  Vogl, Majewski, and Levy}}]{seidl96}
\bibinfo{author}{\bibfnamefont{A.}~\bibnamefont{Seidl}},
  \bibinfo{author}{\bibfnamefont{A.}~\bibnamefont{G{\"o}rling}},
  \bibinfo{author}{\bibfnamefont{P.}~\bibnamefont{Vogl}},
  \bibinfo{author}{\bibfnamefont{J.~A.} \bibnamefont{Majewski}},
  \bibnamefont{and} \bibinfo{author}{\bibfnamefont{M.}~\bibnamefont{Levy}},
  \bibinfo{journal}{Phys.~Rev.~B} \textbf{\bibinfo{volume}{53}},
  \bibinfo{pages}{3764} (\bibinfo{year}{1996}).

\bibitem[{\citenamefont{Becke}(1988)}]{becke88a}
\bibinfo{author}{\bibfnamefont{A.~D.} \bibnamefont{Becke}},
  \bibinfo{journal}{Phys.~Rev.~A} \textbf{\bibinfo{volume}{38}},
  \bibinfo{pages}{3098} (\bibinfo{year}{1988}).

\bibitem[{\citenamefont{Perdew et~al.}(1996{\natexlab{a}})\citenamefont{Perdew,
  Burke, and Ernzerhof}}]{perdew96}
\bibinfo{author}{\bibfnamefont{J.~P.} \bibnamefont{Perdew}},
  \bibinfo{author}{\bibfnamefont{K.}~\bibnamefont{Burke}}, \bibnamefont{and}
  \bibinfo{author}{\bibfnamefont{M.}~\bibnamefont{Ernzerhof}},
  \bibinfo{journal}{Phys.~Rev.~Lett.} \textbf{\bibinfo{volume}{77}},
  \bibinfo{pages}{3865} (\bibinfo{year}{1996}{\natexlab{a}}).

\bibitem[{\citenamefont{Tao et~al.}(2003)\citenamefont{Tao, Perdew, Staroverov,
  and Scuseria}}]{tao03}
\bibinfo{author}{\bibfnamefont{J.}~\bibnamefont{Tao}},
  \bibinfo{author}{\bibfnamefont{J.~P.} \bibnamefont{Perdew}},
  \bibinfo{author}{\bibfnamefont{V.~N.} \bibnamefont{Staroverov}},
  \bibnamefont{and} \bibinfo{author}{\bibfnamefont{G.~E.}
  \bibnamefont{Scuseria}}, \bibinfo{journal}{Phys.~Rev.~Lett.}
  \textbf{\bibinfo{volume}{91}}, \bibinfo{pages}{146401}
  (\bibinfo{year}{2003}).

\bibitem[{\citenamefont{Zhao and Truhlar}(2008)}]{zhao08}
\bibinfo{author}{\bibfnamefont{Y.}~\bibnamefont{Zhao}} \bibnamefont{and}
  \bibinfo{author}{\bibfnamefont{D.~G.} \bibnamefont{Truhlar}},
  \bibinfo{journal}{Theor.~Chem.~Acc.} \textbf{\bibinfo{volume}{41}},
  \bibinfo{pages}{157} (\bibinfo{year}{2008}).

\bibitem[{\citenamefont{Perdew and Schmidt}(2000)}]{perdew00}
\bibinfo{author}{\bibfnamefont{J.~P.} \bibnamefont{Perdew}} \bibnamefont{and}
  \bibinfo{author}{\bibfnamefont{K.}~\bibnamefont{Schmidt}},
  \bibinfo{journal}{AIP Conf.~Proc.} \textbf{\bibinfo{volume}{577}},
  \bibinfo{pages}{1} (\bibinfo{year}{2000}).

\bibitem[{\citenamefont{Becke}(1993)}]{becke93b}
\bibinfo{author}{\bibfnamefont{A.~D.} \bibnamefont{Becke}},
  \bibinfo{journal}{J.~Chem.~Phys.} \textbf{\bibinfo{volume}{98}},
  \bibinfo{pages}{1372} (\bibinfo{year}{1993}).

\bibitem[{\citenamefont{Perdew et~al.}(1996{\natexlab{b}})\citenamefont{Perdew,
  Ernzerhof, and Burke}}]{perdew96a}
\bibinfo{author}{\bibfnamefont{J.~P.} \bibnamefont{Perdew}},
  \bibinfo{author}{\bibfnamefont{M.}~\bibnamefont{Ernzerhof}},
  \bibnamefont{and} \bibinfo{author}{\bibfnamefont{K.}~\bibnamefont{Burke}},
  \bibinfo{journal}{J.~Chem.~Phys.} \textbf{\bibinfo{volume}{105}},
  \bibinfo{pages}{9982} (\bibinfo{year}{1996}{\natexlab{b}}).

\bibitem[{\citenamefont{Becke}(2005)}]{becke05}
\bibinfo{author}{\bibfnamefont{A.~D.} \bibnamefont{Becke}},
  \bibinfo{journal}{J.~Chem.~Phys.} \textbf{\bibinfo{volume}{122}},
  \bibinfo{pages}{064101} (\bibinfo{year}{2005}).

\bibitem[{\citenamefont{Perdew et~al.}(2008)\citenamefont{Perdew, Staroverov,
  Tao, and Scuseria}}]{perdew08}
\bibinfo{author}{\bibfnamefont{J.~P.} \bibnamefont{Perdew}},
  \bibinfo{author}{\bibfnamefont{V.~N.} \bibnamefont{Staroverov}},
  \bibinfo{author}{\bibfnamefont{J.}~\bibnamefont{Tao}}, \bibnamefont{and}
  \bibinfo{author}{\bibfnamefont{G.~E.} \bibnamefont{Scuseria}},
  \bibinfo{journal}{Phys.~Rev.~A} \textbf{\bibinfo{volume}{78}},
  \bibinfo{pages}{052513} (\bibinfo{year}{2008}).

\bibitem[{\citenamefont{Grimme and Neese}(2007)}]{grimme07}
\bibinfo{author}{\bibfnamefont{S.}~\bibnamefont{Grimme}} \bibnamefont{and}
  \bibinfo{author}{\bibfnamefont{F.}~\bibnamefont{Neese}},
  \bibinfo{journal}{J.~Chem.~Phys.} \textbf{\bibinfo{volume}{127}},
  \bibinfo{pages}{154116} (\bibinfo{year}{2007}).

\bibitem[{\citenamefont{Zhang et~al.}(2009)\citenamefont{Zhang, Xu, and
  Goodard}}]{zhang09}
\bibinfo{author}{\bibfnamefont{Y.}~\bibnamefont{Zhang}},
  \bibinfo{author}{\bibfnamefont{X.}~\bibnamefont{Xu}}, \bibnamefont{and}
  \bibinfo{author}{\bibfnamefont{W.}~\bibnamefont{Goodard}},
  \bibinfo{journal}{Proc.~Nat.~Acad.~Sci.~(USA).}
  \textbf{\bibinfo{volume}{106}}, \bibinfo{pages}{4963} (\bibinfo{year}{2009}).

\bibitem[{\citenamefont{Perdew and Wang}(1992)}]{perdew92}
\bibinfo{author}{\bibfnamefont{J.~P.} \bibnamefont{Perdew}} \bibnamefont{and}
  \bibinfo{author}{\bibfnamefont{Y.}~\bibnamefont{Wang}},
  \bibinfo{journal}{Phys.~Rev.~B} \textbf{\bibinfo{volume}{45}},
  \bibinfo{pages}{13244} (\bibinfo{year}{1992}).

\bibitem[{\citenamefont{Yang et~al.}(2000)\citenamefont{Yang, Zhang, and
  Ayers}}]{yang00}
\bibinfo{author}{\bibfnamefont{W.}~\bibnamefont{Yang}},
  \bibinfo{author}{\bibfnamefont{Y.}~\bibnamefont{Zhang}}, \bibnamefont{and}
  \bibinfo{author}{\bibfnamefont{P.~W.} \bibnamefont{Ayers}},
  \bibinfo{journal}{Phys.~Rev.~Lett.} \textbf{\bibinfo{volume}{84}},
  \bibinfo{pages}{5172} (\bibinfo{year}{2000}).

\bibitem[{\citenamefont{Perdew and Zunger}(1981)}]{perdew81}
\bibinfo{author}{\bibfnamefont{J.~P.} \bibnamefont{Perdew}} \bibnamefont{and}
  \bibinfo{author}{\bibfnamefont{A.}~\bibnamefont{Zunger}},
  \bibinfo{journal}{Phys.~Rev.~B} \textbf{\bibinfo{volume}{23}},
  \bibinfo{pages}{5048} (\bibinfo{year}{1981}).

\bibitem[{\citenamefont{Bao et~al.}(2018)\citenamefont{Bao, Gagliardi, and
  Truhlar}}]{bao18}
\bibinfo{author}{\bibfnamefont{J.~L.} \bibnamefont{Bao}},
  \bibinfo{author}{\bibfnamefont{L.}~\bibnamefont{Gagliardi}},
  \bibnamefont{and} \bibinfo{author}{\bibfnamefont{D.~G.}
  \bibnamefont{Truhlar}}, \bibinfo{journal}{J.~Phys.~Chem.~Lett.}
  \textbf{\bibinfo{volume}{9}}, \bibinfo{pages}{2353} (\bibinfo{year}{2018}).

\bibitem[{\citenamefont{Levy et~al.}(1984)\citenamefont{Levy, Perdew, and
  Sahni}}]{levy84}
\bibinfo{author}{\bibfnamefont{M.}~\bibnamefont{Levy}},
  \bibinfo{author}{\bibfnamefont{J.~P.} \bibnamefont{Perdew}},
  \bibnamefont{and} \bibinfo{author}{\bibfnamefont{V.}~\bibnamefont{Sahni}},
  \bibinfo{journal}{Phys.~Rev.~A} \textbf{\bibinfo{volume}{30}},
  \bibinfo{pages}{2745} (\bibinfo{year}{1984}).

\bibitem[{\citenamefont{Kronik and K{\"u}mmel}(2020)}]{kronik20}
\bibinfo{author}{\bibfnamefont{L.}~\bibnamefont{Kronik}} \bibnamefont{and}
  \bibinfo{author}{\bibfnamefont{S.}~\bibnamefont{K{\"u}mmel}},
  \bibinfo{journal}{Phys.~Chem.~Chem.~Phys.} \textbf{\bibinfo{volume}{22}},
  \bibinfo{pages}{16467} (\bibinfo{year}{2020}).

\bibitem[{\citenamefont{Kronik and K{\"u}mmel}(2018)}]{kronik18}
\bibinfo{author}{\bibfnamefont{L.}~\bibnamefont{Kronik}} \bibnamefont{and}
  \bibinfo{author}{\bibfnamefont{S.}~\bibnamefont{K{\"u}mmel}},
  \bibinfo{journal}{Adv.~Mat.} \textbf{\bibinfo{volume}{30}},
  \bibinfo{pages}{1706560} (\bibinfo{year}{2018}).

\bibitem[{\citenamefont{Baer et~al.}(2010)\citenamefont{Baer, Livshits, and
  Salzner}}]{baer10}
\bibinfo{author}{\bibfnamefont{R.}~\bibnamefont{Baer}},
  \bibinfo{author}{\bibfnamefont{E.}~\bibnamefont{Livshits}}, \bibnamefont{and}
  \bibinfo{author}{\bibfnamefont{U.}~\bibnamefont{Salzner}},
  \bibinfo{journal}{Annu.~Rev.~Phys.~Chem.} \textbf{\bibinfo{volume}{61}},
  \bibinfo{pages}{85} (\bibinfo{year}{2010}).

\bibitem[{\citenamefont{Leininger et~al.}(1997)\citenamefont{Leininger, Stoll,
  Werner, and Savin}}]{leininger97}
\bibinfo{author}{\bibfnamefont{T.}~\bibnamefont{Leininger}},
  \bibinfo{author}{\bibfnamefont{H.}~\bibnamefont{Stoll}},
  \bibinfo{author}{\bibfnamefont{H.~J.} \bibnamefont{Werner}},
  \bibnamefont{and} \bibinfo{author}{\bibfnamefont{A.}~\bibnamefont{Savin}},
  \bibinfo{journal}{Chem.~Phys.~Lett.} \textbf{\bibinfo{volume}{275}},
  \bibinfo{pages}{151} (\bibinfo{year}{1997}).

\bibitem[{\citenamefont{Iikura et~al.}(2001)\citenamefont{Iikura, Tsuneda, and
  Hirao}}]{iikura01}
\bibinfo{author}{\bibfnamefont{H.}~\bibnamefont{Iikura}},
  \bibinfo{author}{\bibfnamefont{T.}~\bibnamefont{Tsuneda}}, \bibnamefont{and}
  \bibinfo{author}{\bibfnamefont{K.}~\bibnamefont{Hirao}},
  \bibinfo{journal}{J.~Chem.~Phys.} \textbf{\bibinfo{volume}{115}},
  \bibinfo{pages}{3540} (\bibinfo{year}{2001}).

\bibitem[{\citenamefont{Tawada et~al.}(2004)\citenamefont{Tawada, Tsuneda,
  Yanagisawa, Yanai, and Hirao}}]{tawada04}
\bibinfo{author}{\bibfnamefont{Y.}~\bibnamefont{Tawada}},
  \bibinfo{author}{\bibfnamefont{T.}~\bibnamefont{Tsuneda}},
  \bibinfo{author}{\bibfnamefont{S.}~\bibnamefont{Yanagisawa}},
  \bibinfo{author}{\bibfnamefont{T.}~\bibnamefont{Yanai}}, \bibnamefont{and}
  \bibinfo{author}{\bibfnamefont{K.}~\bibnamefont{Hirao}},
  \bibinfo{journal}{J.~Chem.~Phys.} \textbf{\bibinfo{volume}{120}},
  \bibinfo{pages}{8425} (\bibinfo{year}{2004}).

\bibitem[{\citenamefont{Yanai et~al.}(2004)\citenamefont{Yanai, Tew, and
  Handy}}]{yanai04}
\bibinfo{author}{\bibfnamefont{T.}~\bibnamefont{Yanai}},
  \bibinfo{author}{\bibfnamefont{D.~P.} \bibnamefont{Tew}}, \bibnamefont{and}
  \bibinfo{author}{\bibfnamefont{N.~C.} \bibnamefont{Handy}},
  \bibinfo{journal}{Chem.~Phys.~Lett.} \textbf{\bibinfo{volume}{393}},
  \bibinfo{pages}{51} (\bibinfo{year}{2004}).

\bibitem[{\citenamefont{Lange et~al.}(2008)\citenamefont{Lange, Rohrdanz, and
  Herbert}}]{lange08}
\bibinfo{author}{\bibfnamefont{A.~W.} \bibnamefont{Lange}},
  \bibinfo{author}{\bibfnamefont{M.~A.} \bibnamefont{Rohrdanz}},
  \bibnamefont{and} \bibinfo{author}{\bibfnamefont{J.~M.}
  \bibnamefont{Herbert}}, \bibinfo{journal}{J.~Phys.~Chem.~B}
  \textbf{\bibinfo{volume}{112}}, \bibinfo{pages}{6304} (\bibinfo{year}{2008}).

\bibitem[{\citenamefont{Salzner and Baer}(2009)}]{salzner09}
\bibinfo{author}{\bibfnamefont{U.}~\bibnamefont{Salzner}} \bibnamefont{and}
  \bibinfo{author}{\bibfnamefont{R.}~\bibnamefont{Baer}},
  \bibinfo{journal}{J.~Chem.~Phys.} \textbf{\bibinfo{volume}{131}},
  \bibinfo{pages}{231101} (\bibinfo{year}{2009}).

\bibitem[{\citenamefont{Livshits and Baer}(2007)}]{livshits07}
\bibinfo{author}{\bibfnamefont{E.}~\bibnamefont{Livshits}} \bibnamefont{and}
  \bibinfo{author}{\bibfnamefont{R.}~\bibnamefont{Baer}},
  \bibinfo{journal}{Phys.~Chem.~Chem.~Phys.} \textbf{\bibinfo{volume}{9}},
  \bibinfo{pages}{2932} (\bibinfo{year}{2007}).

\bibitem[{\citenamefont{Stein et~al.}(2010)\citenamefont{Stein, Eisenberg,
  Kronik, and Baer}}]{stein10}
\bibinfo{author}{\bibfnamefont{T.}~\bibnamefont{Stein}},
  \bibinfo{author}{\bibfnamefont{H.}~\bibnamefont{Eisenberg}},
  \bibinfo{author}{\bibfnamefont{L.}~\bibnamefont{Kronik}}, \bibnamefont{and}
  \bibinfo{author}{\bibfnamefont{R.}~\bibnamefont{Baer}},
  \bibinfo{journal}{Phys.~Rev.~Lett.} \textbf{\bibinfo{volume}{105}},
  \bibinfo{pages}{266802} (\bibinfo{year}{2010}).

\bibitem[{\citenamefont{Borpuzari and Kar}(2017)}]{borpuzari17}
\bibinfo{author}{\bibfnamefont{M.~P.} \bibnamefont{Borpuzari}}
  \bibnamefont{and} \bibinfo{author}{\bibfnamefont{R.}~\bibnamefont{Kar}},
  \bibinfo{journal}{J.~Comput.~Chem.} \textbf{\bibinfo{volume}{38}},
  \bibinfo{pages}{2258} (\bibinfo{year}{2017}).

\bibitem[{\citenamefont{Wang and Zhang}(2018)}]{wang18}
\bibinfo{author}{\bibfnamefont{C.}~\bibnamefont{Wang}} \bibnamefont{and}
  \bibinfo{author}{\bibfnamefont{Q.}~\bibnamefont{Zhang}},
  \bibinfo{journal}{J.~Phys.~Chem.~C} \textbf{\bibinfo{volume}{123}},
  \bibinfo{pages}{4407} (\bibinfo{year}{2018}).

\bibitem[{\citenamefont{Tamblyn et~al.}(2014)\citenamefont{Tamblyn,
  Refaely-Abramson, Neaton, and Kronik}}]{tamblyn14}
\bibinfo{author}{\bibfnamefont{I.}~\bibnamefont{Tamblyn}},
  \bibinfo{author}{\bibfnamefont{S.}~\bibnamefont{Refaely-Abramson}},
  \bibinfo{author}{\bibfnamefont{J.~B.} \bibnamefont{Neaton}},
  \bibnamefont{and} \bibinfo{author}{\bibfnamefont{L.}~\bibnamefont{Kronik}},
  \bibinfo{journal}{J.~Phys.~Chem.~Lett.} \textbf{\bibinfo{volume}{5}},
  \bibinfo{pages}{2734} (\bibinfo{year}{2014}).

\bibitem[{\citenamefont{Ghosal et~al.}(2019)\citenamefont{Ghosal, Mandal, and
  Roy}}]{ghosal19}
\bibinfo{author}{\bibfnamefont{A.}~\bibnamefont{Ghosal}},
  \bibinfo{author}{\bibfnamefont{T.}~\bibnamefont{Mandal}}, \bibnamefont{and}
  \bibinfo{author}{\bibfnamefont{A.~K.} \bibnamefont{Roy}},
  \bibinfo{journal}{J.~Chem.~Phys.} \textbf{\bibinfo{volume}{150}},
  \bibinfo{pages}{064104} (\bibinfo{year}{2019}).

\bibitem[{\citenamefont{Ghosal and Roy}(2016)}]{ghosal16}
\bibinfo{author}{\bibfnamefont{A.}~\bibnamefont{Ghosal}} \bibnamefont{and}
  \bibinfo{author}{\bibfnamefont{A.~K.} \bibnamefont{Roy}},
  \emph{\bibinfo{title}{In Specialist Periodical Reports: Chemical Modelling,
  Applications and Theory;~M.~Springborg and J.-O.~Joswig (Eds.) Vol.~13}}
  (\bibinfo{publisher}{{Royal Society of Chemistry}},
  \bibinfo{address}{{London}}, \bibinfo{year}{2016}).

\bibitem[{\citenamefont{Ghosal et~al.}(2018)\citenamefont{Ghosal, Mandal, and
  Roy}}]{ghosal18}
\bibinfo{author}{\bibfnamefont{A.}~\bibnamefont{Ghosal}},
  \bibinfo{author}{\bibfnamefont{T.}~\bibnamefont{Mandal}}, \bibnamefont{and}
  \bibinfo{author}{\bibfnamefont{A.~K.} \bibnamefont{Roy}},
  \bibinfo{journal}{Int.~J.~Quant.~Chem.} \textbf{\bibinfo{volume}{118}},
  \bibinfo{pages}{e25708} (\bibinfo{year}{2018}).

\bibitem[{\citenamefont{Roy}(2008{\natexlab{a}})}]{roy08}
\bibinfo{author}{\bibfnamefont{A.~K.} \bibnamefont{Roy}},
  \bibinfo{journal}{Int.~J.~Quant.~Chem.} \textbf{\bibinfo{volume}{108}},
  \bibinfo{pages}{837} (\bibinfo{year}{2008}{\natexlab{a}}).

\bibitem[{\citenamefont{Roy}(2008{\natexlab{b}})}]{roy08a}
\bibinfo{author}{\bibfnamefont{A.~K.} \bibnamefont{Roy}},
  \bibinfo{journal}{Chem.~Phys.~Lett.} \textbf{\bibinfo{volume}{461}},
  \bibinfo{pages}{142} (\bibinfo{year}{2008}{\natexlab{b}}).

\bibitem[{\citenamefont{Roy}(2010)}]{roy10}
\bibinfo{author}{\bibfnamefont{A.~K.} \bibnamefont{Roy}},
  \bibinfo{journal}{Trends in Phys.~Chem.} \textbf{\bibinfo{volume}{14}},
  \bibinfo{pages}{27} (\bibinfo{year}{2010}).

\bibitem[{\citenamefont{Roy}(2011)}]{roy11}
\bibinfo{author}{\bibfnamefont{A.~K.} \bibnamefont{Roy}},
  \bibinfo{journal}{J.~Math.~Chem.} \textbf{\bibinfo{volume}{49}},
  \bibinfo{pages}{1687} (\bibinfo{year}{2011}).

\bibitem[{\citenamefont{Roy}(2009)}]{roy09}
\bibinfo{author}{\bibfnamefont{A.~K.} \bibnamefont{Roy}}, in
  \emph{\bibinfo{booktitle}{Handbook of Computational Chemistry Research}},
  edited by \bibinfo{editor}{\bibfnamefont{C.~T.} \bibnamefont{Collett}}
  \bibnamefont{and} \bibinfo{editor}{\bibfnamefont{C.~D.} \bibnamefont{Robson}}
  (\bibinfo{publisher}{{Nova publishers}}, \bibinfo{address}{New York},
  \bibinfo{year}{2009}), pp. \bibinfo{pages}{409--434}.

\bibitem[{\citenamefont{Obara and Saika}(1986)}]{obara86}
\bibinfo{author}{\bibfnamefont{S.}~\bibnamefont{Obara}} \bibnamefont{and}
  \bibinfo{author}{\bibfnamefont{A.}~\bibnamefont{Saika}},
  \bibinfo{journal}{J.~Chem.~Phys.} \textbf{\bibinfo{volume}{84}},
  \bibinfo{pages}{3963} (\bibinfo{year}{1986}).

\bibitem[{\citenamefont{Obara and Saika}(1988)}]{obara88}
\bibinfo{author}{\bibfnamefont{S.}~\bibnamefont{Obara}} \bibnamefont{and}
  \bibinfo{author}{\bibfnamefont{A.}~\bibnamefont{Saika}},
  \bibinfo{journal}{J.~Chem.~Phys.} \textbf{\bibinfo{volume}{89}},
  \bibinfo{pages}{1540} (\bibinfo{year}{1988}).

\bibitem[{\citenamefont{Head-Gordon and Pople}(1988)}]{gordon88}
\bibinfo{author}{\bibfnamefont{M.}~\bibnamefont{Head-Gordon}} \bibnamefont{and}
  \bibinfo{author}{\bibfnamefont{J.~A.} \bibnamefont{Pople}},
  \bibinfo{journal}{J.~Chem.~Phys.} \textbf{\bibinfo{volume}{89}},
  \bibinfo{pages}{5777} (\bibinfo{year}{1988}).

\bibitem[{\citenamefont{Liu et~al.}(2016)\citenamefont{Liu, Furlani, and
  Kong}}]{liu16}
\bibinfo{author}{\bibfnamefont{F.}~\bibnamefont{Liu}},
  \bibinfo{author}{\bibfnamefont{T.}~\bibnamefont{Furlani}}, \bibnamefont{and}
  \bibinfo{author}{\bibfnamefont{J.}~\bibnamefont{Kong}},
  \bibinfo{journal}{J.~Phys.~Chem.~A} \textbf{\bibinfo{volume}{120}},
  \bibinfo{pages}{10264} (\bibinfo{year}{2016}).

\bibitem[{\citenamefont{Baer and Neuhauser}(2005)}]{baer05}
\bibinfo{author}{\bibfnamefont{R.}~\bibnamefont{Baer}} \bibnamefont{and}
  \bibinfo{author}{\bibfnamefont{D.}~\bibnamefont{Neuhauser}},
  \bibinfo{journal}{Phys.~Rev.~Lett.} \textbf{\bibinfo{volume}{94}},
  \bibinfo{pages}{043002} (\bibinfo{year}{2005}).

\bibitem[{\citenamefont{Mandal et~al.}(2019)\citenamefont{Mandal, Ghosal, and
  Roy}}]{mandal19}
\bibinfo{author}{\bibfnamefont{T.}~\bibnamefont{Mandal}},
  \bibinfo{author}{\bibfnamefont{A.}~\bibnamefont{Ghosal}}, \bibnamefont{and}
  \bibinfo{author}{\bibfnamefont{A.~K.} \bibnamefont{Roy}},
  \bibinfo{journal}{Theor.~Chem.~Acc.} \textbf{\bibinfo{volume}{138}},
  \bibinfo{pages}{10} (\bibinfo{year}{2019}).

\bibitem[{\citenamefont{K{\"o}rzd{\"o}rfer
  et~al.}(2011)\citenamefont{K{\"o}rzd{\"o}rfer, Sears, Sutton, and
  Br{\'e}das}}]{korzdorfer11}
\bibinfo{author}{\bibfnamefont{T.}~\bibnamefont{K{\"o}rzd{\"o}rfer}},
  \bibinfo{author}{\bibfnamefont{J.~S.} \bibnamefont{Sears}},
  \bibinfo{author}{\bibfnamefont{C.}~\bibnamefont{Sutton}}, \bibnamefont{and}
  \bibinfo{author}{\bibfnamefont{J.-L.} \bibnamefont{Br{\'e}das}},
  \bibinfo{journal}{J.~Chem.~Phys.} \textbf{\bibinfo{volume}{135}},
  \bibinfo{pages}{204107} (\bibinfo{year}{2011}).

\bibitem[{\citenamefont{Martyna and Tuckerman}(1999)}]{martyna99}
\bibinfo{author}{\bibfnamefont{G.~J.} \bibnamefont{Martyna}} \bibnamefont{and}
  \bibinfo{author}{\bibfnamefont{M.~E.} \bibnamefont{Tuckerman}},
  \bibinfo{journal}{J.~Chem.~Phys.} \textbf{\bibinfo{volume}{110}},
  \bibinfo{pages}{2810} (\bibinfo{year}{1999}).

\bibitem[{\citenamefont{Kronik et~al.}(2012)\citenamefont{Kronik, Stein,
  Refaely-Abramson, and Baer}}]{kronik12}
\bibinfo{author}{\bibfnamefont{L.}~\bibnamefont{Kronik}},
  \bibinfo{author}{\bibfnamefont{T.}~\bibnamefont{Stein}},
  \bibinfo{author}{\bibfnamefont{S.}~\bibnamefont{Refaely-Abramson}},
  \bibnamefont{and} \bibinfo{author}{\bibfnamefont{R.}~\bibnamefont{Baer}},
  \bibinfo{journal}{J.~Chem.~Theory~Comput.} \textbf{\bibinfo{volume}{8}},
  \bibinfo{pages}{1515} (\bibinfo{year}{2012}).

\bibitem[{\citenamefont{Rohrdanz et~al.}(2009)\citenamefont{Rohrdanz, Martins,
  and Herbert}}]{rohrdanz09}
\bibinfo{author}{\bibfnamefont{M.~A.} \bibnamefont{Rohrdanz}},
  \bibinfo{author}{\bibfnamefont{K.~M.} \bibnamefont{Martins}},
  \bibnamefont{and} \bibinfo{author}{\bibfnamefont{J.~M.}
  \bibnamefont{Herbert}}, \bibinfo{journal}{J.~Chem.~Phys.}
  \textbf{\bibinfo{volume}{130}}, \bibinfo{pages}{054112}
  (\bibinfo{year}{2009}).

\bibitem[{\citenamefont{Chai and Head-Gordon}(2008{\natexlab{a}})}]{chai08a}
\bibinfo{author}{\bibfnamefont{J.-D.} \bibnamefont{Chai}} \bibnamefont{and}
  \bibinfo{author}{\bibfnamefont{M.}~\bibnamefont{Head-Gordon}},
  \bibinfo{journal}{Chem.~Phys.~Lett.} \textbf{\bibinfo{volume}{467}},
  \bibinfo{pages}{176} (\bibinfo{year}{2008}{\natexlab{a}}).

\bibitem[{\citenamefont{Chai and Head-Gordon}(2008{\natexlab{b}})}]{chai08b}
\bibinfo{author}{\bibfnamefont{J.-D.} \bibnamefont{Chai}} \bibnamefont{and}
  \bibinfo{author}{\bibfnamefont{M.}~\bibnamefont{Head-Gordon}},
  \bibinfo{journal}{J.~Chem.~Phys.} \textbf{\bibinfo{volume}{128}},
  \bibinfo{pages}{084106} (\bibinfo{year}{2008}{\natexlab{b}}).

\bibitem[{\citenamefont{Peverati and Truhlar}(2012)}]{peverati12}
\bibinfo{author}{\bibfnamefont{R.}~\bibnamefont{Peverati}} \bibnamefont{and}
  \bibinfo{author}{\bibfnamefont{D.~G.} \bibnamefont{Truhlar}},
  \bibinfo{journal}{Phys.~Chem.~Chem,.~Phys.} \textbf{\bibinfo{volume}{14}},
  \bibinfo{pages}{16187} (\bibinfo{year}{2012}).

\bibitem[{\citenamefont{Vikramaditya et~al.}(2018)\citenamefont{Vikramaditya,
  Chai, and Lin}}]{vikramaditya18}
\bibinfo{author}{\bibfnamefont{T.}~\bibnamefont{Vikramaditya}},
  \bibinfo{author}{\bibfnamefont{J.-D.} \bibnamefont{Chai}}, \bibnamefont{and}
  \bibinfo{author}{\bibfnamefont{S.-T.} \bibnamefont{Lin}},
  \bibinfo{journal}{J.~Comp.~Chem.} \textbf{\bibinfo{volume}{39}},
  \bibinfo{pages}{2378} (\bibinfo{year}{2018}).

\bibitem[{\citenamefont{Chan et~al.}(2019)\citenamefont{Chan, Kawashima, and
  Hirao}}]{chan19}
\bibinfo{author}{\bibfnamefont{B.}~\bibnamefont{Chan}},
  \bibinfo{author}{\bibfnamefont{Y.}~\bibnamefont{Kawashima}},
  \bibnamefont{and} \bibinfo{author}{\bibfnamefont{K.}~\bibnamefont{Hirao}},
  \bibinfo{journal}{J.~Comput.~Chem.} \textbf{\bibinfo{volume}{40}},
  \bibinfo{pages}{29} (\bibinfo{year}{2019}).

\bibitem[{\citenamefont{Henderson et~al.}(2008)\citenamefont{Henderson,
  Janesko, and Scuseria}}]{henderson08}
\bibinfo{author}{\bibfnamefont{T.~M.} \bibnamefont{Henderson}},
  \bibinfo{author}{\bibfnamefont{B.~G.} \bibnamefont{Janesko}},
  \bibnamefont{and} \bibinfo{author}{\bibfnamefont{G.}~\bibnamefont{Scuseria}},
  \bibinfo{journal}{J.~Chem.~Phys.} \textbf{\bibinfo{volume}{128}},
  \bibinfo{pages}{194105} (\bibinfo{year}{2008}).

\bibitem[{\citenamefont{Cohen et~al.}(2007)\citenamefont{Cohen,
  Mori-S{\'a}nchez, and Yang}}]{cohen07}
\bibinfo{author}{\bibfnamefont{A.~J.} \bibnamefont{Cohen}},
  \bibinfo{author}{\bibfnamefont{P.}~\bibnamefont{Mori-S{\'a}nchez}},
  \bibnamefont{and} \bibinfo{author}{\bibfnamefont{W.}~\bibnamefont{Yang}},
  \bibinfo{journal}{J.~Chem.~Phys.} \textbf{\bibinfo{volume}{126}},
  \bibinfo{pages}{191109} (\bibinfo{year}{2007}).

\bibitem[{\citenamefont{Lin et~al.}(2012)\citenamefont{Lin, Tsai, Li, and
  Chai}}]{lin12}
\bibinfo{author}{\bibfnamefont{Y.-S.} \bibnamefont{Lin}},
  \bibinfo{author}{\bibfnamefont{C.-W.} \bibnamefont{Tsai}},
  \bibinfo{author}{\bibfnamefont{G.-D.} \bibnamefont{Li}}, \bibnamefont{and}
  \bibinfo{author}{\bibfnamefont{J.-D.} \bibnamefont{Chai}},
  \bibinfo{journal}{J.~Chem.~Phys.} \textbf{\bibinfo{volume}{136}},
  \bibinfo{pages}{154109} (\bibinfo{year}{2012}).

\bibitem[{\citenamefont{Johnson et~al.}(1993)\citenamefont{Johnson, Gill, and
  Pople}}]{johnson93}
\bibinfo{author}{\bibfnamefont{B.~G.} \bibnamefont{Johnson}},
  \bibinfo{author}{\bibfnamefont{P.~M.~W.} \bibnamefont{Gill}},
  \bibnamefont{and} \bibinfo{author}{\bibfnamefont{J.~A.} \bibnamefont{Pople}},
  \bibinfo{journal}{J.~Chem.~Phys.} \textbf{\bibinfo{volume}{98}},
  \bibinfo{pages}{5612} (\bibinfo{year}{1993}).

\bibitem[{\citenamefont{Ernzerhof and Perdew}(1998)}]{ernzerhof98}
\bibinfo{author}{\bibfnamefont{M.}~\bibnamefont{Ernzerhof}} \bibnamefont{and}
  \bibinfo{author}{\bibfnamefont{J.~P.} \bibnamefont{Perdew}},
  \bibinfo{journal}{J.~Chem.~Phys.} \textbf{\bibinfo{volume}{109}},
  \bibinfo{pages}{3313} (\bibinfo{year}{1998}).

\bibitem[{\citenamefont{{A.~K.~Roy, A.~Ghosal and T.~Mandal}}(2019)}]{indft19}
\bibinfo{author}{\bibnamefont{{A.~K.~Roy, A.~Ghosal and T.~Mandal}}},
  \emph{\bibinfo{title}{InDFT: A DFT Program for Atoms and Molecules in CCG}}
  (\bibinfo{publisher}{Theoretical Chemistry Laboratory},
  \bibinfo{address}{{IISER Kolkata, India}}, \bibinfo{year}{2019}),
  \bibinfo{note}{{This is based on the extension of an initial version of the
  code, established by A. K. Roy, in 2008, whose results were published in
  refs. [41-45]}}.

\bibitem[{\citenamefont{Stephens et~al.}(1994)\citenamefont{Stephens, Devlin,
  Chabalowski, and Frisch}}]{stephens94}
\bibinfo{author}{\bibfnamefont{P.~J.} \bibnamefont{Stephens}},
  \bibinfo{author}{\bibfnamefont{F.~J.} \bibnamefont{Devlin}},
  \bibinfo{author}{\bibfnamefont{C.~F.} \bibnamefont{Chabalowski}},
  \bibnamefont{and} \bibinfo{author}{\bibfnamefont{M.~J.}
  \bibnamefont{Frisch}}, \bibinfo{journal}{J.~Chem.~Phys.}
  \textbf{\bibinfo{volume}{98}}, \bibinfo{pages}{11623} (\bibinfo{year}{1994}).

\bibitem[{\citenamefont{Schmidt et~al.}(1993)\citenamefont{Schmidt, Baldridge,
  Boatz, Elbert, Gordon, Hensen, Koseki, Matsunaga, Nguyen, Su
  et~al.}}]{schmidt93}
\bibinfo{author}{\bibfnamefont{M.~W.} \bibnamefont{Schmidt}},
  \bibinfo{author}{\bibfnamefont{K.~K.} \bibnamefont{Baldridge}},
  \bibinfo{author}{\bibfnamefont{J.~A.} \bibnamefont{Boatz}},
  \bibinfo{author}{\bibfnamefont{S.~T.} \bibnamefont{Elbert}},
  \bibinfo{author}{\bibfnamefont{M.~S.} \bibnamefont{Gordon}},
  \bibinfo{author}{\bibfnamefont{J.~H.} \bibnamefont{Hensen}},
  \bibinfo{author}{\bibfnamefont{S.}~\bibnamefont{Koseki}},
  \bibinfo{author}{\bibfnamefont{N.}~\bibnamefont{Matsunaga}},
  \bibinfo{author}{\bibfnamefont{K.~A.} \bibnamefont{Nguyen}},
  \bibinfo{author}{\bibfnamefont{S.~J.} \bibnamefont{Su}},
  \bibnamefont{et~al.}, \bibinfo{journal}{J.~Comput.~Chem.}
  \textbf{\bibinfo{volume}{14}}, \bibinfo{pages}{1347} (\bibinfo{year}{1993}).

\bibitem[{\citenamefont{Stevens et~al.}(1984)\citenamefont{Stevens, Basch, and
  Krauss}}]{stevens84}
\bibinfo{author}{\bibfnamefont{W.~J.} \bibnamefont{Stevens}},
  \bibinfo{author}{\bibfnamefont{H.}~\bibnamefont{Basch}}, \bibnamefont{and}
  \bibinfo{author}{\bibfnamefont{M.}~\bibnamefont{Krauss}},
  \bibinfo{journal}{J.~Chem.~Phys.} \textbf{\bibinfo{volume}{81}},
  \bibinfo{pages}{6026} (\bibinfo{year}{1984}).

\bibitem[{\citenamefont{Hay and Wadt}(1985)}]{hay85c}
\bibinfo{author}{\bibfnamefont{P.~J.} \bibnamefont{Hay}} \bibnamefont{and}
  \bibinfo{author}{\bibfnamefont{W.~R.} \bibnamefont{Wadt}},
  \bibinfo{journal}{J.~Chem.~Phys.} \textbf{\bibinfo{volume}{82}},
  \bibinfo{pages}{284} (\bibinfo{year}{1985}).

\bibitem[{\citenamefont{Feller}(1996)}]{feller96}
\bibinfo{author}{\bibfnamefont{D.}~\bibnamefont{Feller}},
  \bibinfo{journal}{J.~Comp.~Chem.} \textbf{\bibinfo{volume}{17}},
  \bibinfo{pages}{1571} (\bibinfo{year}{1996}).

\bibitem[{\citenamefont{Vosko et~al.}(1980)\citenamefont{Vosko, Wilk, and
  Nusair}}]{vosko80}
\bibinfo{author}{\bibfnamefont{S.~H.} \bibnamefont{Vosko}},
  \bibinfo{author}{\bibfnamefont{L.}~\bibnamefont{Wilk}}, \bibnamefont{and}
  \bibinfo{author}{\bibfnamefont{M.}~\bibnamefont{Nusair}},
  \bibinfo{journal}{Can.~J.~Phys.} \textbf{\bibinfo{volume}{58}},
  \bibinfo{pages}{1200} (\bibinfo{year}{1980}).

\bibitem[{\citenamefont{Lee et~al.}(1988)\citenamefont{Lee, Yang, and
  Parr}}]{lee88}
\bibinfo{author}{\bibfnamefont{C.}~\bibnamefont{Lee}},
  \bibinfo{author}{\bibfnamefont{W.}~\bibnamefont{Yang}}, \bibnamefont{and}
  \bibinfo{author}{\bibfnamefont{R.~G.} \bibnamefont{Parr}},
  \bibinfo{journal}{Phys.~Rev.~B} \textbf{\bibinfo{volume}{37}},
  \bibinfo{pages}{785} (\bibinfo{year}{1988}).

\bibitem[{rep(2001)}]{repository}
\emph{\bibinfo{title}{Density Functional Repository, Quantum Chemistry Group}}
  (\bibinfo{publisher}{CCLRC Daresbury Laboratory},
  \bibinfo{address}{{Daresbury, Cheshire, UK}}, \bibinfo{year}{2001}).

\bibitem[{\citenamefont{Frigo and Johnson}(2005)}]{fftw05}
\bibinfo{author}{\bibfnamefont{M.}~\bibnamefont{Frigo}} \bibnamefont{and}
  \bibinfo{author}{\bibfnamefont{S.~G.} \bibnamefont{Johnson}},
  \bibinfo{journal}{Proceedings of the IEEE} \textbf{\bibinfo{volume}{93}},
  \bibinfo{pages}{216} (\bibinfo{year}{2005}).

\bibitem[{\citenamefont{Anderson et~al.}(1999)\citenamefont{Anderson, Bai,
  Bischof, Blackford, Dongarra, Greenbaum, Hammarling, A.~McKenney, and
  Sorensen}}]{anderson99}
\bibinfo{author}{\bibfnamefont{E.}~\bibnamefont{Anderson}},
  \bibinfo{author}{\bibfnamefont{Z.}~\bibnamefont{Bai}},
  \bibinfo{author}{\bibfnamefont{C.}~\bibnamefont{Bischof}},
  \bibinfo{author}{\bibfnamefont{S.}~\bibnamefont{Blackford}},
  \bibinfo{author}{\bibfnamefont{J.}~\bibnamefont{Dongarra}},
  \bibinfo{author}{\bibfnamefont{J.~D.~C.~A.} \bibnamefont{Greenbaum}},
  \bibinfo{author}{\bibfnamefont{S.}~\bibnamefont{Hammarling}},
  \bibinfo{author}{\bibfnamefont{A.}~\bibnamefont{A.~McKenney}},
  \bibnamefont{and} \bibinfo{author}{\bibfnamefont{D.}~\bibnamefont{Sorensen}},
  \emph{\bibinfo{title}{LAPACK Users' Guide}}, vol.~\bibinfo{volume}{9}
  (\bibinfo{publisher}{SIAM}, \bibinfo{year}{1999}).

\bibitem[{\citenamefont{Johnson}(2016)}]{johnson16}
\bibinfo{author}{\bibfnamefont{R.~D.} \bibnamefont{Johnson}},
  \emph{\bibinfo{title}{III~(Ed.) NIST Computational Chemistry Comparisons and
  Benchmark Database, NIST Standard Reference Database, Number, Release 18}}
  (\bibinfo{publisher}{NIST}, \bibinfo{address}{{Gaithersburg, MD}},
  \bibinfo{year}{2016}).

\bibitem[{\citenamefont{Ghosal et~al.}(2021)\citenamefont{Ghosal, Gupta,
  Mahato, and Roy}}]{ghosal21}
\bibinfo{author}{\bibfnamefont{A.}~\bibnamefont{Ghosal}},
  \bibinfo{author}{\bibfnamefont{T.}~\bibnamefont{Gupta}},
  \bibinfo{author}{\bibfnamefont{K.}~\bibnamefont{Mahato}}, \bibnamefont{and}
  \bibinfo{author}{\bibfnamefont{A.~K.} \bibnamefont{Roy}},
  \bibinfo{journal}{Theor. Chem. Acc.} \textbf{\bibinfo{volume}{140}},
  \bibinfo{pages}{1} (\bibinfo{year}{2021}).

\bibitem[{\citenamefont{Perdew et~al.}(1982)\citenamefont{Perdew, Parr, Levy,
  and Balduz~Jr{.}}}]{perdew82}
\bibinfo{author}{\bibfnamefont{J.~P.} \bibnamefont{Perdew}},
  \bibinfo{author}{\bibfnamefont{R.~G.} \bibnamefont{Parr}},
  \bibinfo{author}{\bibfnamefont{M.}~\bibnamefont{Levy}}, \bibnamefont{and}
  \bibinfo{author}{\bibfnamefont{J.~L.} \bibnamefont{Balduz~Jr{.}}},
  \bibinfo{journal}{Phys.~Rev.~Lett.} \textbf{\bibinfo{volume}{49}},
  \bibinfo{pages}{1691} (\bibinfo{year}{1982}).

\bibitem[{\citenamefont{Perdew and Levy}(1997)}]{perdew97}
\bibinfo{author}{\bibfnamefont{J.~P.} \bibnamefont{Perdew}} \bibnamefont{and}
  \bibinfo{author}{\bibfnamefont{M.}~\bibnamefont{Levy}},
  \bibinfo{journal}{Phys.~Rev.~B} \textbf{\bibinfo{volume}{56}},
  \bibinfo{pages}{16021} (\bibinfo{year}{1997}).

\bibitem[{\citenamefont{G{\"o}rling and Levy}(1997)}]{gorling97}
\bibinfo{author}{\bibfnamefont{A.}~\bibnamefont{G{\"o}rling}} \bibnamefont{and}
  \bibinfo{author}{\bibfnamefont{M.}~\bibnamefont{Levy}},
  \bibinfo{journal}{J.~Chem.~Phys.} \textbf{\bibinfo{volume}{106}},
  \bibinfo{pages}{2675} (\bibinfo{year}{1997}).

\bibitem[{\citenamefont{Pople et~al.}(1989)\citenamefont{Pople, Head-Gordon,
  Fox, Raghavachari, and Curtiss}}]{pople89}
\bibinfo{author}{\bibfnamefont{J.~A.} \bibnamefont{Pople}},
  \bibinfo{author}{\bibfnamefont{M.}~\bibnamefont{Head-Gordon}},
  \bibinfo{author}{\bibfnamefont{D.~J.} \bibnamefont{Fox}},
  \bibinfo{author}{\bibfnamefont{K.}~\bibnamefont{Raghavachari}},
  \bibnamefont{and} \bibinfo{author}{\bibfnamefont{L.~A.}
  \bibnamefont{Curtiss}}, \bibinfo{journal}{J.~Chem.~Phys.}
  \textbf{\bibinfo{volume}{90}}, \bibinfo{pages}{5622} (\bibinfo{year}{1989}).

\bibitem[{\citenamefont{Landau}(1957)}]{landau57}
\bibinfo{author}{\bibfnamefont{L.~D.} \bibnamefont{Landau}},
  \bibinfo{journal}{Sov.~Phys.~JETP} \textbf{\bibinfo{volume}{3}},
  \bibinfo{pages}{920} (\bibinfo{year}{1957}).

\bibitem[{\citenamefont{Mori-S{\'a}nchez
  et~al.}(2006)\citenamefont{Mori-S{\'a}nchez, Cohen, and Yang}}]{mori06}
\bibinfo{author}{\bibfnamefont{P.}~\bibnamefont{Mori-S{\'a}nchez}},
  \bibinfo{author}{\bibfnamefont{A.~J.} \bibnamefont{Cohen}}, \bibnamefont{and}
  \bibinfo{author}{\bibfnamefont{W.}~\bibnamefont{Yang}},
  \bibinfo{journal}{J.~Chem.~Phys.} \textbf{\bibinfo{volume}{124}},
  \bibinfo{pages}{091102} (\bibinfo{year}{2006}).

\bibitem[{\citenamefont{Janak}(1978)}]{janak78}
\bibinfo{author}{\bibfnamefont{J.~F.} \bibnamefont{Janak}},
  \bibinfo{journal}{Phys.~Rev.~B} \textbf{\bibinfo{volume}{18}},
  \bibinfo{pages}{7165} (\bibinfo{year}{1978}).

\end{thebibliography}

\end{document}